\documentclass[final,5p,times,twocolumn,sort&compress]{elsarticle}
\usepackage{xr} 
\usepackage[colorlinks=true,linkcolor=blue,citecolor=blue,urlcolor=blue]{hyperref}

\usepackage[utf8]{inputenc}
\usepackage{graphicx}
\usepackage{dcolumn}
\usepackage{bm}
\usepackage{longtable}
\usepackage{lscape}
\usepackage{multirow}
\usepackage{booktabs}
\usepackage{amsmath}
\usepackage{float}
\usepackage{rotating}
\usepackage{caption}
\usepackage{placeins}
\usepackage{xcolor}
\usepackage{miller}
\usepackage{gensymb}
\usepackage{algorithm}
\usepackage{algpseudocode}
\usepackage{acronym}

\acrodef{ACE}{atomic cluster expansion}
\acrodef{EAM}{embedded atom method}
\acrodef{DFT}{density functional theory}
\acrodef{VASP}{vienna ab-initio simulation package}
\acrodef{NEB}{nudged elastic band}
\acrodef{MEP}{minimum energy path}
\acrodef{FIRE}{fast inertial relaxation engine}
\acrodef{LAMMPS}{large-scale atomic/molecular massively parallel simulator}
\acrodef{DXA}{dislocation extraction algorithm}
\acrodef{OVITO}{open visualization tool}
\acrodef{DBSCAN}{density-based spatial clustering of applications with noise}
\acrodef{LaCA}{local atomic cluster analysis}
\acrodef{HAADF-STEM}{high-angle annular dark-field scanning transmission electron microscopy}
\acrodef{FIB}{focused ion beam}
\acrodef{TEM}{transmission electron microscopy}
\acrodef{ISF}{intrinsic stacking fault}
\acrodef{ESF}{extrinsic stacking fault}
\acrodef{SFE}{stacking fault energy}
\acrodef{TB}{twin boundary}
\acrodef{RC}{reaction coordinate}
\acrodef{ESE}{excess strain energy}
\acrodef{TCP}{topologically close-packed}
\acrodef{FCC}{face-centered cubic}
\acrodef{HCP}{hexagonal close-packed}
\acrodef{BCC}{body-centered cubic}
\acrodef{BDTT}{brittle-to-ductile transition temperature}

\usepackage{graphicx}
\usepackage{subcaption}
\usepackage[percent]{overpic}
\usepackage[inline]{showlabels}

\begin{document}

\begin{frontmatter}

\title{Zonal dislocations in Laves phases: A coupled synchro-shear slip mechanism}

\author[1]{Sang-Hyeok Lee}
\affiliation[1]{organization={Institut für Metallkunde und Materialphysik, RWTH Aachen University},
    city={52056 Aachen},
    country={Germany}}

\author[2]{Mariano Forti}
\affiliation[2]{organization={Interdisciplinary Centre for Advanced Materials Simulation (ICAMS), Ruhr University Bochum, Universitätsstrasse 150},
    city={44801 Bochum},
    country={Germany}}

\author[2]{Thomas Hammerschmidt\corref{cor1}}
\cortext[cor1]{Corresponding author}
\ead{thomas.hammerschmidt@rub.de}

\author[3,4]{Gang Liu\corref{cor1}}
\affiliation[3]{organization={Max Planck Institute for Sustainable Materials},
    city={40237 Düsseldorf},
    country={Germany}}
\affiliation[4]{organization={School of Materials Science and Engineering, Fuyao University of Science and Technology},
    city={Fuzhou, Fujian 350109},
    country={China}}
\ead{g.liu@fyust.edu.cn}

\author[5]{Julien Guénolé}
\affiliation[5]{organization={Université de Lorraine, CNRS, Arts et Métiers, LEM3},
    city={57070 Metz},
    country={France}}
    
\author[3,6]{Siyuan Zhang}
\affiliation[6]{organization={Helmholtz-Zentrum Berlin für Materialien und Energie},
    city={12489 Berlin},
    country={Germany}}

\author[3]{Gerhard Dehm}

\author[1]{Sandra Korte-Kerzel}

\author[1]{Zhuocheng Xie\corref{cor1}}
\ead{xie@imm.rwth-aachen.de}

\begin{abstract}
Synchro-shear is the primary plastic deformation mechanism in Laves phases at elevated temperatures, mediated by synchro-Shockley partial dislocations—zonal dislocations that proceed via localized events such as kink-pair nucleation and propagation. Using atomistic simulations, we identified a novel slip mechanism in Laves phases, namely coupled synchro-shear slip, involving the synchronized glide of two synchro-Shockley partial dislocations on adjacent slip planes, leading to the formation of extrinsic stacking faults. High-resolution scanning transmission electron microscopy revealed extended core structures consistent with coupled synchro-Shockley partial dislocations bounded by extrinsic stacking faults in \textit{C}15 NbCr$_{2}$ and their involvement in twinning. These results highlight the critical role of coupled synchro-shear slip in enabling phase transformations between Laves polytypes and in governing twinning behavior, providing new atomistic insight into the kinetic nature of plasticity in topologically close-packed intermetallic phases.

\end{abstract}

\begin{keyword}
Laves phase; zonal dislocation; stacking fault; atomistic simulation; HAADF-STEM

\end{keyword}

\end{frontmatter}

\section{Introduction}

Zonal dislocations, first introduced to describe plastic deformation in complex crystal structures such as sapphire \cite{kronberg1957plastic}, represent a unique class of dislocations characterized by the correlated glide of multiple partial dislocations with different Burgers vectors on adjacent atomic planes \cite{anderson2017theory}. This collective motion enables shear and atomic shuffling to occur simultaneously across neighboring layers, thereby accommodating plastic deformation in crystals with intricate stacking sequences \cite{kazantzis1997deformation,kishida2022direct,mussi2023experimental,stollenwerk2024dislocation}. The concept of zonal dislocations has also been extended to describe twinning mechanisms in hexagonal close-packed metals, where a single twin dislocation shears only part of the matrix into the twin orientation, while the remaining atoms achieve the correct twin orientation through atomic shuffling \cite{bilby1965theory,mendelson1969zonal,wang20091}. 

\begin{figure*}[hbt!]
\centering
\includegraphics[width=\textwidth]{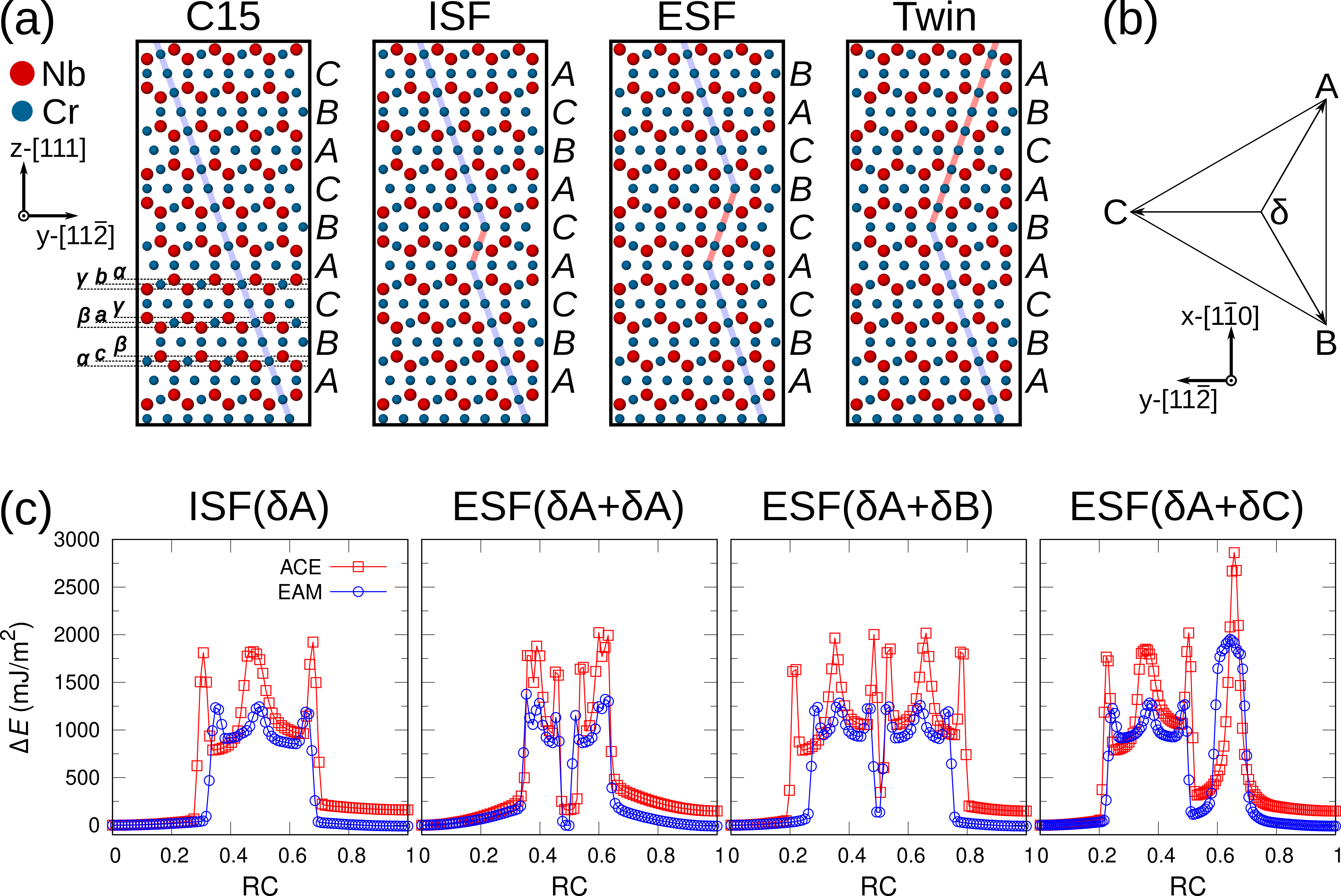}
\caption{Planar fault formation in \textit{C}15 Laves phases. (a) Crystal and (111) planar fault structures in \textit{C}15 Laves phases. Red (large) and blue (small) spheres represent Nb and Cr atoms, respectively. (b) Thompson tetrahedron notation of (111) slip systems in \textit{C}15 Laves phases. (c) The minimal energy paths for (111) planar fault formation in the \textit{C}15 NbCr$_2$ Laves phase calculated using the EAM\_Rösch and ACE potentials: intrinsic stacking fault formation via one partial slip event ($\delta A$) and extrinsic stacking fault formation via two partial slip events ($\delta A + \delta A/\delta B/\delta C$) on adjacent triple layers.}
\label{fig:fault}
\end{figure*}

Zonal dislocations provide a fundamental framework for understanding plastic deformation in structurally complex intermetallic systems such as Laves phases. Plastic deformation on the \hkl{111} or basal planes in Laves phases is primarily governed by the synchro-shear slip mechanism \cite{2005science,stein2021laves}. The corresponding dislocation, known as a synchro-Shockley dislocation, is a type of zonal dislocation that involves the synchronized glide of two coupled Shockley partial dislocations on adjacent planes within a triple-layer unit. In \textit{C}15 (or \textit{C}14) Laves phases, the slip of synchro-Shockley dislocations transforms alternating triple layers into slabs of \textit{C}14 (or \textit{C}15) structure, thereby forming intrinsic stacking faults (ISFs) (see Figure~\ref{fig:fault}(a)). This mechanism is believed to play a central role in polytypic transformations between Laves phases and in twinning within \textit{C}15 structures \cite{livingston1990room,chu1993deformation,kumar1996laves}.

The synchro-shear slip mechanism and its minimum energy path (MEP) in Laves phases have been investigated extensively using ab-initio and atomistic simulations \cite{vedmedenko2008first,guenole2019basal,ladines2023off}. Recent atomistic simulations have provided new insights into the motion mechanisms of synchro-Shockley partial dislocations associated with the extension of ISFs \cite{xie2023unveiling}. Two distinct types of 30\degree\ synchro-Shockley partials have been identified, each exhibiting a different mode of motion: type I propagates via kink-pair nucleation and propagation, while type II moves through a three-stage process involving non-sequential atomic shuffling, shear straining, and atomic rearrangement. The motion of synchro-Shockley dislocations is thermally activated, with thermal assistance being essential to overcoming shear-insensitive events, implying that their motion is suppressed at low temperatures \cite{xie2023thermally}. 

While these findings have deepened our understanding of ISF-related processes, experimental observations have revealed more complex planar faults in Laves phases \cite{allen1972electron,liao1979crystal,liu1995defect,luzzi1997analysis,luzzi1998deformation,chu1998phase,wang2024revealing,vslapakova2020atomic,zhang2020shuffle,xie2024temperature,zhang2025dissociation}. Among these, \hkl{111} extrinsic stacking faults (ESFs), which can be regarded as the superposition of two ISFs (see Figure~\ref{fig:fault}(a)), have been frequently reported following heat treatment or deformation \cite{allen1972electron,liao1979crystal,liu1995defect,luzzi1997analysis,luzzi1998deformation,chu1998phase,wang2024revealing}. In several \textit{C}15 Laves phases, ESFs are found to be more prevalent than ISFs \cite{allen1972electron,chu1998phase} due to their lower formation energies \cite{hong1999elastic,hong2000first,sun2004ab,yao2007first}. However, the dislocation-mediated mechanisms governing the formation of ESFs remain unclear. Furthermore, previous studies have proposed that twinning in \textit{C}15 Laves phases may occur through a sequential formation of ISFs on adjacent \hkl{111} planes, analogous to the classical twinning mechanism in face-centered cubic (FCC) metals, achieved by the consecutive motion of synchro-Shockley dislocations \cite{livingston1990room,chu1993deformation,kumar1996laves}. Yet, it remains unclear whether the proposed twinning mechanism is energetically favorable or kinetically feasible in all Laves phases.

Among all Laves phases, \textit{C}15 NbCr$_{2}$ is one of the most extensively studied systems, particularly regarding its deformation behavior above the brittle-to-ductile transition temperature (BDTT) \cite{vignoul1990high,takasugi1995high,yoshida1994deformation,kazantzis1996stacking,kazantzis1997deformation,yoshida2002tem,kazantzis2007mechanical,kazantzis2008self}. Deformation twinning has been observed in high-temperature-deformed \textit{C}15 NbCr$_{2}$, exhibiting an anomalous dependence on both strain rate and temperature compared with FCC metals \cite{kazantzis2007mechanical}. Microstructural analyses further revealed that twinning preferentially occurs in grains oriented to activate two co-planar partial slip systems with comparably high resolved shear stresses \cite{kazantzis2007mechanical}. In addition, synchro-Shockley partials in \textit{C}15 NbCr$_{2}$ exhibit a self-pinning behavior, where dislocation velocity lags behind changes in strain rate, implying that mobile dislocation density evolves through thermally activated motion \cite{kazantzis2008self}. Alloying additions of V \cite{vignoul1993characterization}, Mo \cite{takasugi1996deformability}, or Y \cite{lu2009fracture} in \textit{C}15 NbCr$_{2}$ have been shown to induce solid-solution softening and enhance fracture toughness. Notably, \textit{C}15 NbCr$_{2}$ with 5\,at.\% V shows no evidence of deformation twinning; instead, plastic deformation is accommodated primarily by the glide and climb of full dislocations \cite{yoshida2003tem}.

In this study, the mechanisms of dislocation motion in the \textit{C}15 NbCr$_{2}$ Laves phase were elucidated using atomistic simulations. The results provide new insights into the formation of ESFs, which are mediated by the coordinated glide of two coupled synchro-Shockley partial dislocations on adjacent planes, collectively referred to as a zonal dislocation, as revealed by nudged elastic band (NEB) calculations. Furthermore, an exhaustive high-throughput brute-force approach was performed to screen partial dislocation pairs that are likely to glide together (attractive partials) and those that prefer to glide separately (repulsive partials). In addition, high-angle annular dark-field scanning transmission electron microscopy (HAADF-STEM) confirmed the presence of coupled synchro-Shockley partial dislocations bounded by ESFs in a NbCr$_2$ Laves phase after high-temperature annealing. Finally, a new twinning mechanism involving consecutive motion of coupled synchro-Shockley partial dislocations was unveiled. These findings establish a direct link between atomistic dislocation mechanisms and experimentally observed defect structures, providing a comprehensive understanding of zonal dislocation behavior in Laves phases.

\section{Methods}

\subsection{Interatomic potentials}

\subsubsection{Development of ACE potential}

Interatomic interactions were modeled using an atomic cluster expansion potential (ACE) \cite{ace_drautz2019}, which was parametrized using the \textit{pacemaker} package \cite{lysogorskiy2021performant, bochkarev2022efficient}. The ACE model and the underlying dataset of density functional theory (DFT) calculations will be presented in detail elsewhere.

\paragraph{Reference dataset}
A reference dataset was built to train the ACE potential.
Besides BCC, FCC and HCP-based structures (both ordered and special quasirandom structures), the Laves phases \textit{C}14, \textit{C}15 and \textit{C}36 were also included. 
In addition, other topologically close-packed (TCP) phases were considered to ensure the inclusion of all relevant interatomic environments: \textit{A}15, $\sigma$ and $\mu$. 
For all of them, all possible configurations are considered by permutation of Cr and Nb in all Wyckoff sites, as in typical sublattice model approximations.
In order to train the potential in dislocation-like environments, \textit{C}15 generalized stacking fault structures are also included. 
Lastly, the potential is prepared to perform well during relaxation in finite temperature conditions by including randomly deformed structures in the training set.
In total, the dataset is formed by approximately 44,800 structures with 861,000 atoms.

\paragraph{DFT calculations}
Reference energies and forces were calculated for each structure in the reference dataset using DFT as implemented in the \textit{Vienna ab-initio simulation package} (VASP) \cite{kresse1993ab, kresse1994ab, kresse1996efficiency}. 
The plane wave basis expansion of the wave functions was set up with a cutoff energy of 500\,eV.
Electronic occupations were determined with a Gaussian smearing with a width of 0.1\,eV.
Brillouin zone integration was performed with a Monkhorst-Pack \cite{Monkhorst1976} grid with a spacing of 0.125\,$\AA^{-1}$.
The generalized gradient approximation (GGA) to the Exchange-correlation functional was used in the Perdew-Burke-Ernzerhof (PBE)\cite{Perdew1996} parametrization.
Structure optimization was performed with a force tolerance of 5\,meV/$\AA$, while the self-consistent energy cycle was considered converged with an energy tolerance of 10\textsuperscript{-6}\,eV.

\paragraph{Parameterization of the ACE potential}
The ACE model takes into account up to 5-body interactions, within a cutoff radius of 7\,$\AA$. 
Core repulsion interactions are switched on at 1.5\,$\AA$.
The final potential uses a total of 5,648 basis functions with a total of 16,788 trained parameters. 

\subsubsection{Potential benchmark}
Two alternative semi-empirical embedded atom method (EAM) interatomic potentials, EAM\_Rösch \cite{rosch2006interatomic} and EAM\_Heaton \cite{heaton2024first}, developed for the \textit{C}15 NbCr\textsubscript{2} Laves phase, were also used in this study. Our ACE and EAM\_Rösch potentials demonstrated good accuracy in predicting lattice parameters, formation energies, and elastic constants compared to experimental data and DFT calculations (see Table \ref{tab:potential-benchmark}).

\begin{table*}[!htbp]
\centering
\caption[]{\label{tab:potential-benchmark}Properties of \textit{C}15 NbCr\textsubscript{2} Laves phase calculated using the EAM and ACE potentials. $a\textsubscript{0}$: lattice parameters; $E_\text{f}$: formation energy; $C\textsubscript{ij}$: elastic constants; ISF energy: \hkl{111} intrinsic stacking fault energy; ESF energy: \hkl{111} extrinsic stacking fault energy; TB energy: \hkl{111} twin boundary energy.}
\centering
\begin{tabular}{l@{\hspace{1cm}}c@{\hspace{1cm}}c@{\hspace{1cm}}c@{\hspace{1cm}}c@{\hspace{1cm}}c@{\hspace{1cm}}}
\hline\hline
\addlinespace[0.1cm]
Properties &  EAM\_Rösch & EAM\_Heaton & ACE & DFT \\
\addlinespace[0.1cm]
\hline
\addlinespace[0.1cm]
$a\textsubscript{0}$ (\text{\AA}) & 6.939 & 6.807 & 6.948 & 6.931~\cite{yao2006first}, 6.948~\cite{long2016predicting}, 6.958~\cite{liu2017first} \\
$E_\text{f}^{\textit{C}14}$ (eV/atom) & -0.053 & -0.389 & -0.036 & -0.006~\cite{hong1999phase}  \\
$E_\text{f}^{\textit{C}15}$ (eV/atom) & -0.052 & -0.381 & -0.055 & -0.026~\cite{hong1999phase}, -0.072~\cite{yao2006first},-0.073~\cite{ormeci1996total} \\
$E_\text{f}^{\textit{C}36}$ (eV/atom) & -0.053 & -0.385 & -0.047 & -0.019~\cite{hong1999phase}  \\
$C\textsubscript{11}$ (GPa) & 298.8 & 350.5 & 341.4 & 298~\cite{liu2017first}, 316~\cite{hong1999phase}, 309~\cite{rosch2006interatomic} \\
$C\textsubscript{12}$ (GPa) & 180.5 & 148.4 & 184.5 & 191~\cite{liu2017first}, 216~\cite{hong1999phase}, 198~\cite{rosch2006interatomic} \\
$C\textsubscript{44}$ (GPa) & 55.5 & 91.5 & 73.4 & 65~\cite{liu2017first}, 71~\cite{hong1999phase}, 69~\cite{rosch2006interatomic} \\
ISF energy\,(mJ/m$^{2}$) & -10.3 & -77.7 & 161.1 & 116~\cite{hong2000first}, 128~\cite{ma2014ab,vedmedenko2008first}, 170~\cite{liu2017first}, 180~\cite{chu1995stacking} \\
ESF energy\,(mJ/m$^{2}$) & -10.2 & -79.3 & 148.8 & 91~\cite{ma2014ab}, 94~\cite{hong2000first} \\
TB energy\,(mJ/m$^{2}$) & -5.2 & -35.9 & 74.4 & 39~\cite{hong2000first}, 52~\cite{vedmedenko2008first} \\
\hline\hline
\end{tabular}
\end{table*}

\subsection{Atomistic configurations}
\subsubsection{Simulation cell}
The \textit{C}15 NbCr$_{2}$ Laves phase structures were generated using Atomsk \cite{hirel2015atomsk} with the following crystallographic orientations: $\mathbf{x}\parallel BA=\hkl[1-10]$, $\mathbf{y} \parallel C\delta =\hkl[11-2]$ and $\mathbf{z} \parallel D\delta =\hkl[111]$. For the stacking fault energy (SFE) calculations and NEB calculations on rigid body displacement, the simulation cell contains $1 \times 1 \times 15$ unit cells with periodic boundary conditions in the $\mathbf{x}$ and $\mathbf{y}$ directions. Stacking fault and dislocation structures were characterized using LaCA \cite{xie2021laves} to identify atoms satisfying TCP packing criteria.

\subsubsection{High-throughput dislocation screening}
The configurational space of individual synchro-Shockley dislocations and pairs of synchro-Shockley dislocations on adjacent slip planes was systematically explored by varying the core positions of individual partials (Figure \ref{fig:core}) and arithmetically hybridizing the displacement of two of them (Figure \ref{fig:paircore}) to yield ESF configurations with systematically varied $\mathbf{y}$-position space.

A pristine \textit{C}15 NbCr$_{2}$ slab with dimensions $l_x$= 0.98\,nm, $l_y$= 93.6\,nm, and $l_z$= 90.2\,nm was first constructed. A single synchro-Shockley partial dislocation bounding an ISF was introduced, with periodic boundary conditions applied along the line direction, $\boldsymbol{\xi}\parallel\mathbf{x}=\hkl[1-10]$, and gliding on the \hkl(111) plane. All symmetry equivalent Burgers vector orientations of the $\frac{1}{6}\hkl<112>$ type, generated at $60\degree$ intervals, were sampled. The $\mathbf{y}$ and $\mathbf{z}$ insertion positions were additionally varied to identify the metastable dislocation core configurations and to sample equivalent cores located in neighboring Peierls valleys and adjacent atomic layers (cf. Figure \ref{fig:core}). The resulting dislocation configurations were subsequently relaxed using the FIRE algorithm \cite{bitzek2006structural,guenole2020assessment} with a force tolerance of $10^{-8}$\,eV/$\AA$. The procedure yielded four energetically preferable unique single partial configurations: $30\degree$ and $90\degree$ type I and type II synchro-Shockley partials. The relaxed dislocation positions and core structures for all six directions are given in Figures~\ref{fig:si-ab-npos}--\ref{fig:si-cb-npos} of the Supplementary Information (SI). For a pair of synchro-Shockley partial dislocations, the $\mathbf{y}$ coordinates of one of the core positions were systematically varied to sample consecutive Peierls valleys (cf. Figure \ref{fig:paircore}).

After relaxation, the dislocation core positions were determined using the dislocation extraction algorithm (DXA) implemented in OVITO \cite{stukowski2009visualization, stukowski2012automated} on the A-atom (Nb) sublattice.
The excess strain energy of the dislocation core ($E_{ESE}$) was evaluated from the excess potential energy relative to the cohesive energies of Nb and Cr atoms in perfect \textit{C}15 NbCr$_{2}$. 
Cylindrical regions centered on the identified dislocation core position were defined, and the total excess strain energy, $E_{ESE}$, was summed as a function of the cylinder radius $r$, see Figure~\ref{fig:core}. 
Rather than prescribing a single SFE for all configurations, an effective stacking-fault correction was determined separately for each configuration by maximizing the linearity of the corrected $E_{ESE}$ with respect to $\ln(r/b)$ in the far-field region, where $b=|\mathbf{b}|$ is the magnitude of the Burgers vector. This procedure provides a core-agnostic correction for the stacking fault contribution with multiple participating stacking faults and is therefore suitable for automated high-throughput analysis.
After applying the stacking fault correction, the resulting excess energies were fitted linearly as a function of $\ln(r/b)$. The fitted energy at $r/b=1$, corresponding to $\ln(r/b)=0$, was taken as the dislocation core energy.

The relaxed single-dislocation configurations were clustered using a DBSCAN \cite{ester1996density} clustering process aiming to exclude high-energy metastable dislocation configurations. The core positions were clustered along the slip direction $\mathbf{y}$ ($\varepsilon=2.5$, min\_sample = 1) and then only the lowest-energy configuration within each positional cluster was taken. The screening results are shown in Figure~\ref{fig:si-screening} of the SI.
The core energies of the two-dislocation configurations were evaluated using the same excess strain energy method. For each pair, an effective position was defined by averaging the $\mathbf{y}$ and $\mathbf{z}$ coordinates of the two constituent dislocation cores.

\subsubsection{Nanopillar}
For the surface dislocation nucleation calculations, a \textit{C}15 NbCr$_{2}$ cylinder with a diameter of 6.1\,nm and height of 18.3\,nm was constructed with and without an ESF, oriented such that the \hkl(111) slip plane formed a 45\degree\ inclination with respect to the pillar axis. The structures were relaxed using the FIRE algorithm with a force tolerance of $10^{-8}$\,eV/$\AA$.

\subsection{Nudged elastic band calculations}
Atomistic simulations were performed using LAMMPS \cite{LAMMPS} in this study. The NEB method \cite{henkelman2000climbing,henkelman2000improved} was used to investigate the mechanisms of dislocation motion and the corresponding energy profiles. The initial configuration was created by inserting a dislocation at the center of a pristine simulation cell, as described above. A final configuration was then generated by introducing an identical dislocation at the adjacent Peierls valley, offset by one or a few minimum translational symmetry distances. To ensure the MEP primarily reflected dislocation motion, discrepancies in surface structures were eliminated by imposing the far-field structure of the final configuration onto the initial configuration, ensuring consistent boundary conditions. The simulation cell dimensions were set to $l_y$=37.3\,nm and $l_z$=36.1\,nm. Periodic boundary conditions were applied along the line direction in $\mathbf{x}$ with dimensions $l_x$ of 0.98\,nm and 5.89\,nm. These systems contained approximately 95,000 and 570,000 atoms, respectively. Semi-fixed boundary conditions were applied to the outermost layers of the configurations with a thickness of more than twice the interatomic potential cutoff. All dislocation structures were minimized using conjugate gradient and FIRE algorithms with a force tolerance of $10^{-8}$\,eV/$\AA$. The spring constants for parallel and perpendicular nudging forces were set to 1.0\,eV/$\AA^{2}$ \cite{MARAS201613}. The Quickmin algorithm \cite{sheppard2008optimization} was used as the damped dynamics minimizer to minimize the energies across all replicas with a force tolerance of 0.01\,eV/$\AA$. Different numbers of intermediate replicas from 48 to 192 were simulated and all intermediate replicas were initially equally spaced along the reaction coordinate (RC).

\subsection{Experiments}
Off-stoichiometric Cr-35.0\,at.\% Nb alloys were synthesized as cylindrical ingots weighing approximately 350 grams, with a diameter of 15\,mm and a length of approximately 100\,mm, through levitation melting in a cold crucible system. The starting materials consisted of chromium with 99.95\,wt.\% purity and niobium with 99.9\,wt.\% purity. The molten alloy was subsequently drop-cast into an alumina crucible that had been preheated to 1200\,\degree C. The temperature was maintained for 45 minutes, followed by cooling of the crucible at a controlled rate of 5\,K/min to ambient temperature. Further, a homogenization heat treatment was conducted at 1200\,\degree C for 96~hours under a high-purity argon atmosphere, and then the alloy was slowly cooled at a rate of 5\,K/min to room temperature. 
Focused ion beam (FIB)  lift-out (FEI Scios 2) was used to extract a site-specific lamella for further transmission electron microscopy (TEM) analysis. Atomistic characterization of the TEM sample was conducted by HAADF-STEM (probe-corrected Titan Themis from Thermo Fisher Scientific, operated at 300\,kV). For HAADF-STEM images, the probe convergence semi-angle was 23.8\,mrad and the collection semi-angles were 78-200\,mrad.

\section{Results}
\subsection{Planar fault formation in \textit{C}15 NbCr$_{2}$}\label{result:fault}

Typical \hkl{111} planar fault structures in the \textit{C}15 NbCr$_{2}$ Laves phase are shown in Figure~\ref{fig:fault}(a). In the \textit{C}15 structure, the lattice is composed of repeating quadruple units, where Cr atomic layers in the Kagomé and triple layers are labeled with uppercase and lowercase Roman letters, respectively, while Nb atomic layers in the triple layers are labeled with lowercase Greek letters. The stacking sequence along the \hkl[111] direction is therefore expressed as ...$A\alpha c\beta B\beta a\gamma C\gamma b\alpha$..., as illustrated in Figure~\ref{fig:fault}(a). Several types of planar defects, ISF, ESF, and twin, are also presented using this notation in Figure~\ref{fig:fault}(a). Neglecting the triple layers for simplicity, the stacking sequence of an ISF along the \hkl[111] direction is ...$ABCA\mid CABCA$..., where a $B$ atomic layer is missing. Similarly, an ESF can be described as ...$ABCA\underline{C}BCAB$..., with the underlined $C$ atomic layer representing an additional stacking layer. A twin exhibits the sequence ...$ABCA\underline{B}ACBA$..., with the underlined $B$ marking the twin boundary, while the \textit{C}36 structure can be interpreted as a periodic repetition of ESFs separated by two quadruple units (not visualized in Figure~\ref{fig:fault}(a)).

The MEPs for the formation of ISFs and ESFs in the \textit{C}15 NbCr$_{2}$ Laves phase were calculated using both the EAM\_Rösch and ACE interatomic potentials (Figure~\ref{fig:fault}(c)). The formation of an ISF proceeds through partial slip events along one of three equivalent $\frac{1}{6}\hkl<112>$ slip directions on the \hkl(111) plane, denoted as $\delta A$, $\delta B$, and $\delta C$ in the Thompson tetrahedron notation (Figure~\ref{fig:fault}(b)). The slip mechanism follows a synchro-shear process, in which the energy barrier decomposes into three sequential local maxima corresponding to shear and atomic shuffling events.
The formation of an ESF can be viewed as a combination of two consecutive ISF formation events, yielding three possible partial slip combinations: $\delta A + \delta A$, $\delta A + \delta B$, and $\delta A + \delta C$. The associated energy profiles similarly decompose into sequential ISF formation stages (Figure~\ref{fig:fault}(c)). Notably, the slip event along $\delta C$ does not exhibit clear dissociation into distinct peaks, suggesting a different degree of coupling between shear and shuffling in the $1 \times 1 \times 15$ unit cells. In all cases, the EAM\_Rösch potential systematically underestimates the energy barriers compared to the ACE potential, yet both predict identical energy profile shapes and atomic-scale mechanisms, demonstrating consistent physical fidelity.

\subsection{Core structures and energies of zonal dislocations in \textit{C}15 NbCr$_{2}$}\label{result:core}

\begin{figure*}[hbt!]
    \centering
\includegraphics[
        width=\linewidth,
        height=0.90\textheight,
        keepaspectratio]{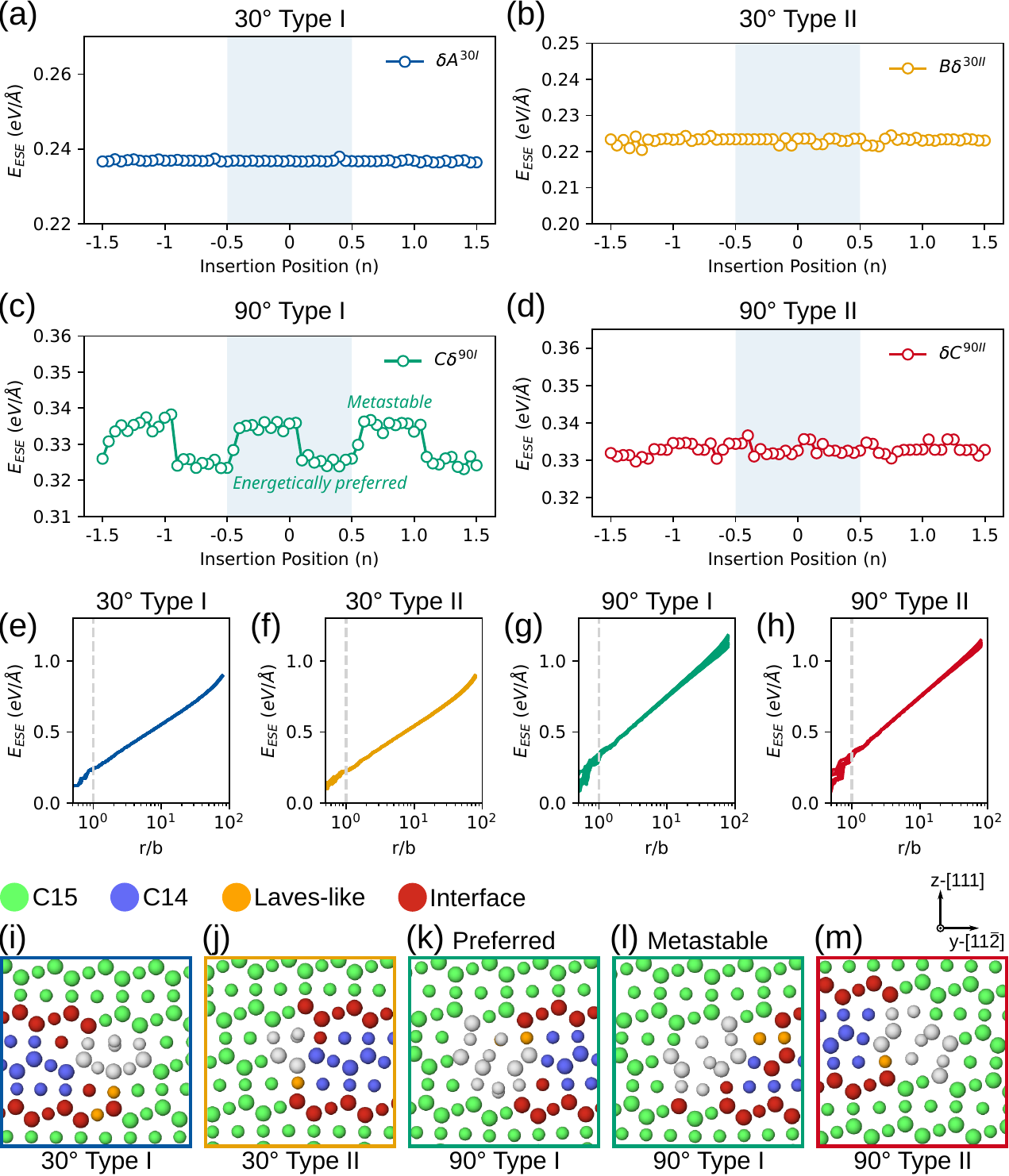}
\caption{Screening on core structures and energies of synchro-Shockley partial dislocations with the line direction $\boldsymbol{\xi}\parallel\mathbf{x}=\hkl[1-10]$ in \textit{C}15 NbCr$_{2}$ by varying the intended insertion $\mathbf{y}$ position: (a) $30\degree$ type~I, (b) $30\degree$ type~II, (c) $90\degree$ type~I, and (d) $90\degree$ type~II. The excess strain energies of preferred dislocation cores: (e) $30\degree$ type~I, (f) $30\degree$ type~II, (g) $90\degree$ type~I, and (h) $90\degree$ type~II. Unique dislocation configurations: (i) $30\degree$ type~I, (j) $30\degree$ type II, (k) energetically preferred $90\degree$ type I, (l) metastable $90\degree$ type I, and (m) $90\degree$ type II. Atoms are colored according to their structural motif using LaCA\cite{xie2021laves}.}
\label{fig:core}
\end{figure*}

\begin{figure*}[hbt!]
\centering
    \includegraphics[width=0.9\linewidth]{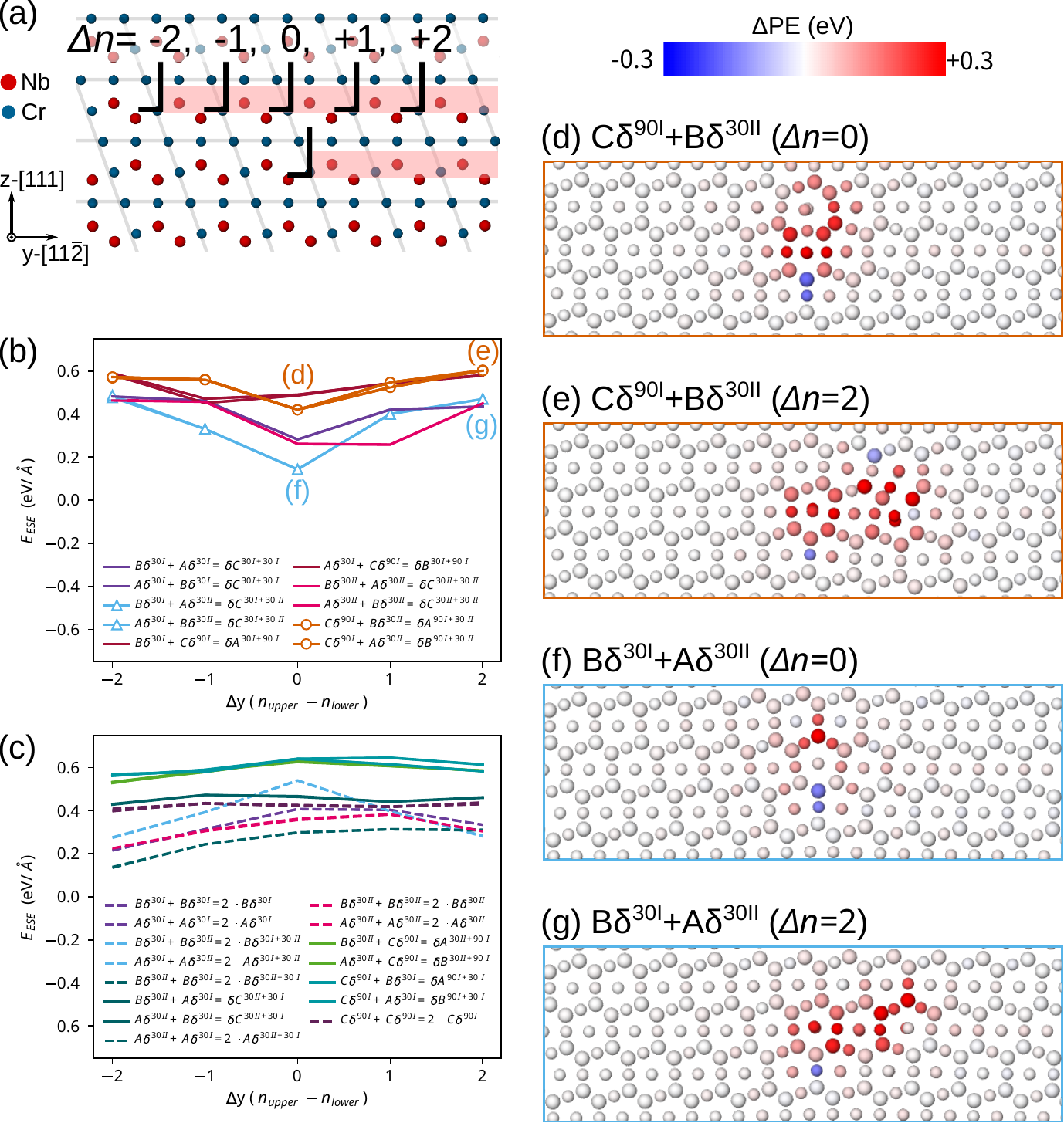}
\caption{Energy profiles of synchro-Shockley partial dislocation pairs as a function of their relative separation along the glide direction. (a) Schematic illustration of the five relative positions of a pair of synchro-Shockley partial dislocations across consecutive Peierls valleys ($\Delta n=-2, -1, 0, +1, +2$). Excess strain energy profiles of pairs of synchro-Shockley partial dislocations with the line direction $\boldsymbol{\xi}\parallel\mathbf{x}=\hkl[1-10]$, evaluated as a function of their relative positions across five consecutive Peierls valleys on adjacent slip planes, are shown in panel (b) for the attractive and panel (c) for the repulsive cases. Dislocations are labeled using Roman-Greek notation, with the superscript specifying the character angle followed by the core-geometry type. Pairs with identical Burgers vectors are represented by dashed lines. The pair combinations similar to the experimental observations are circled in (b). Core structures of the $C\delta^{90\text{I}}+B\delta^{30\text{II}}$ pair at $\Delta n=0$ and $\Delta n=+2$ are shown in (d) and (e), respectively, and those of the $B\delta^{30\text{I}}+A\delta^{30\text{II}}$ pair at $\Delta n=0$ and $\Delta n=+2$ are shown in (f) and (g), respectively. Atoms are colored according to their excess potential energies ($\Delta$PE).}
\label{fig:paircore}
\end{figure*}

Partial dislocations gliding on the \hkl{111} plane via the synchro-shear slip mechanism, which leads to the formation of ISFs, are referred to as synchro-Shockley partial dislocations. Each synchro-Shockley partial dislocation consists of two coupled partial slip components: one associated with shear between the triple-Kagomé layers and the other involving local atomic shuffling within the triple layer. Owing to this cooperative character, the synchro-Shockley partial dislocation can be regarded as a zonal dislocation, distinguished by the coordinated motion of dislocations on parallel adjacent slip planes. The naming convention and characterization of the synchro-Shockley partial dislocations in this study are given in Table \ref{tab:coretype}. For simplification, the Thompson tetrahedron label of the Burgers vector is assigned from the displacement of the A sub-lattice using the spatial Burgers vector extracted using OVITO. This Burgers vector determines the far-field elastic strain. However, it does not uniquely specify the microscopic core structure, which additionally depends on the Burgers vector of the second partial slip component responsible for the local atomic shuffling. Although this component modifies the core geometry, it produces no net far-field strain. Each synchro-Shockley partial is therefore denoted using a label such as $A\delta^{30II}$. The Roman--Greek or Greek--Roman symbols identify the magnitude, direction, and sense of the net Burgers vector according to the Thompson-tetrahedron notation. The superscript specifies the dislocation character, defined by the angle between the net Burgers vector and the dislocation-line direction, followed by the core-geometry variant. Accordingly, $30\mathrm{I}$, $30\mathrm{II}$, $90\mathrm{I}$, and $90\mathrm{II}$ denote the $30\degree$ type~I, $30\degree$ type~II, $90\degree$ type~I, and $90\degree$ type~II synchro-Shockley partials, respectively. 

\begin{table}[htbp]
    \centering
    \small
    \caption{Initial dislocation insertion parameters (full dislocation Burgers vector $\mathrm{b}$ and line vector $\mathrm{\xi}$) and the resulting left (Roman--Greek) and right (Greek--Roman) partials, with their partial Burgers vector and dislocation core geometry type after relaxation. The sample coordinate system is aligned with the Thompson tetrahedron such that $+\mathbf{x}\parallel{BA}$, $+\mathbf{y}\parallel {\delta C}$, and $+\mathbf{z}\parallel D\delta$.}
    \label{tab:coretype}
    \renewcommand{\arraystretch}{1.25}
    \begin{tabular}{cc|cccc}
        \toprule
         \multicolumn{2}{c}{\textbf{Full dislocation}}
          & \multicolumn{2}{c}{\textbf{Left partial}}
          & \multicolumn{2}{c}{\textbf{Right partial}} \\
          $\mathbf{b}$ & $\mathbf{\xi}$ & $\mathbf{b}$ & Core
              & $\mathbf{b}$ & Core \\
        \midrule
          $BA$ & $AB$ & $B\delta$  & $30\degree$ Type II & $\delta A$ & $30\degree$ Type I \\
          $BC$ & $AB$ & $B\delta$  & $30\degree$ Type I  & $\delta C$ & $90\degree$ Type II \\
          $AC$ & $AB$ & $A\delta$  & $30\degree$ Type I  & $\delta C$ & $90\degree$ Type II \\
          $AB$ & $AB$ & $A\delta$  & $30\degree$ Type II & $\delta B$ & $30\degree$ Type I \\
          $CB$ & $AB$ & $C\delta$  & $90\degree$ Type I  & $\delta B$ & $30\degree$ Type II \\
          $CA$ & $AB$ & $C\delta$  & $90\degree$ Type I  & $\delta A$ & $30\degree$ Type II \\
        \bottomrule
    \end{tabular}
\end{table}

For a dislocation line direction along \hkl[1-10] ($AB$), four types with different Burgers vector orientations were examined both energetically and geometrically: $30\degree$ type I and type II, and $90\degree$ type I and type II. Their calculated metastable and preferred core structures and energies are shown in Figure~\ref{fig:core}. 
The $90\degree$ partials have substantially higher core energies than the $30\degree$ partials. The two partial types also differ in core structure metastability: the $30\degree$ partials show no metastable configurations (Figure \ref{fig:core}(i–j)), while the $90\degree$ type I partials exhibit energetically distinguishable metastable states (Figure \ref{fig:core}(k-l)).
Two synchro-Shockley dislocations gliding on adjacent slip planes lead to the formation of an ESF. The energy profiles of the two synchro-Shockley dislocations (both Roman-Greek) were constructed by varying the relative positions across five consecutive Peierls valleys, see Figure \ref{fig:paircore}(a). One partial was positioned at $n=-2,-1,0,+1,+2$ relative to the other partial, which was kept fixed. Here, $n=0$ corresponds to zero lateral motif offset, such that the two cores are aligned on adjacent slip planes. Because the configurations differ only in their relative core positions, their total core energies can be compared directly to assess the thermodynamic preference for spatial association.
The dislocation pairs were classified into two characteristic forms, attractive and repulsive, according to the type of discrete turning points in these energy profiles (Figure \ref{fig:paircore}(b-c)). A configuration at $n=i$, where $i=-1,0,+1$, was identified as a local minimum when its energy was lower than those of both neighboring configurations; a local maximum was defined analogously. Profiles containing a significant turning point corresponding to a local minimum were classified as attractive, indicating an energetically preferred relative position at which the two partials tend to remain spatially coupled. Conversely, profiles containing a significant turning point corresponding to a local maximum were classified as repulsive, indicating that the corresponding close pair configuration was energetically unfavorable and that increased separation was preferred.

As shown in Figure~\ref{fig:paircore}, all dislocation pairs with identical Burgers vectors exhibited pronounced repulsive behavior, with energy maxima located at one of the interior positions, $n=-1,0,+1$, or exhibited monotonic increase or decrease behavior. In contrast, most pairs with different Burgers vectors exhibited attractive behavior, characterized by an interior energy minimum. Combinations of Greek--Roman partials are given in Figure~\ref{fig:greek-roman-esf} of the SI. This classification describes the thermodynamic preference for spatial association but does not establish whether the two partials preferentially glide cooperatively or independently. Such a judgment requires exploration of the minimum energy paths, which is evaluated through further NEB calculations in the next section.

\begin{figure*}[hbt!]
\centering
\includegraphics[width=\linewidth]{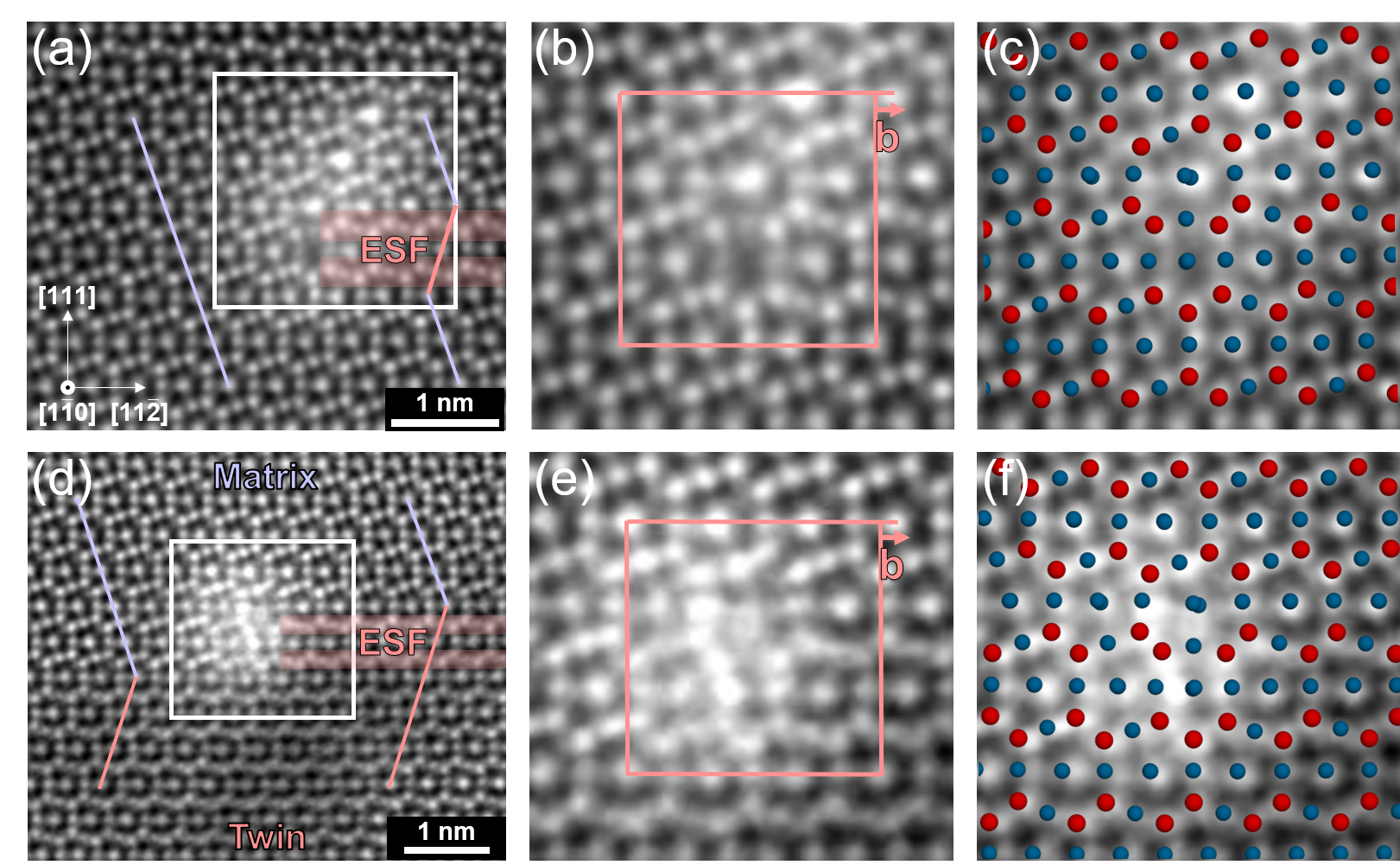}
\caption{Core structures of zonal dislocations bounded by ESFs propagating (a) in the matrix and (d) along a coherent twin boundary, representing twinning. (b,e) Zoomed-in views of the dislocation core regions in (a) and (d) with Burgers circuits imposed on the images. (c,f) Atomistic simulation configurations of the dislocation core structures overlaid with (b) and (e). Red (large) and blue (small) spheres represent Nb and Cr atoms, respectively.}
\label{fig:exp_sim}
\end{figure*}

ESFs and pairs of synchro-Shockley dislocations gliding on adjacent \hkl{111} planes were observed by high-resolution HAADF-STEM, as shown in Figure~\ref{fig:exp_sim}. Figure \ref{fig:exp_sim}(a) presents an ESF propagating in the \textit{C}15 matrix, where two synchro-Shockley dislocations can be treated as a single zonal dislocation. A Burgers circuit constructed around the dislocation core region revealed a closure failure corresponding to a partial Burgers vector of $\delta A$ or $\delta B$ (projected as $\frac{C \delta}{2}$), which can be interpreted as the combination of a 90\degree\ type I synchro-Shockley partial $C\delta$ ($\frac{1}{6}\hkl[-1-12]$) and a 30\degree\ type II synchro-Shockley partial, either $A \delta$ ($\frac{1}{6}\hkl[-12-1]$) or $B \delta$ ($\frac{1}{6}\hkl[2-1-1]$). Atomistic simulations were performed to model this dislocation configuration, in which a 90\degree\ type I and a 30\degree\ type II synchro-Shockley partial were introduced at the upper and lower triple layers, respectively, within the \textit{C}15 matrix. The simulated dislocation core configuration closely reproduced the experimentally observed structure, as shown in Figure \ref{fig:exp_sim}(c). Furthermore, the $90\degree$ + $30\degree$ dislocation combinations, e.g., $C\delta^{90I}+B\delta^{30II}$ and $C\delta^{90I}+A\delta^{30II}$, exhibited similar energy profiles across the sampled Peierls valleys, each featuring a pronounced local minimum (see Figure~\ref{fig:paircore}). Both pairs were therefore classified as attractive, indicating a thermodynamic preference for spatial association.

Figure \ref{fig:exp_sim}(d) shows an ESF propagating along a coherent twin boundary, representing a twinning-related configuration. The core structure of the pair of synchro-Shockley partial dislocations bounding the ESF is shown in Figure \ref{fig:exp_sim}(e), where Burgers circuit closure failure was identified along the \hkl(111) plane. The zonal dislocation exhibits the same overall Burgers vector, with a total projected component of $\frac{C\delta}{2}$ or $\frac{1}{12}\hkl[-1-12]$, consistent with a combined 90\degree\ type I and 30\degree\ type II synchro-Shockley partial pair, as observed in the \textit{C}15 matrix. Atomistic simulations introducing this coupled dislocation configuration adjacent to a coherent twin boundary reproduce the experimentally observed structure (Figure~\ref{fig:exp_sim}(f)).

Additionally, a pronounced increase in HAADF intensity at the dislocation cores (Figure~\ref{fig:exp_sim}) and a decrease in the lower region indicate local Nb imbalance, likely arising from diffusion-controlled segregation during annealing or cooling. This enrichment induces local lattice distortion, which may account for the slight deviations between the experimental and simulated dislocation configurations, the latter being modeled under stoichiometric conditions. To isolate the intrinsic dislocation motion mechanisms from solute effects, point defect decoration at the dislocation core was deliberately excluded from the following calculations.

\begin{figure*}[hbpt!]
    \centering
    \includegraphics[width=\linewidth]{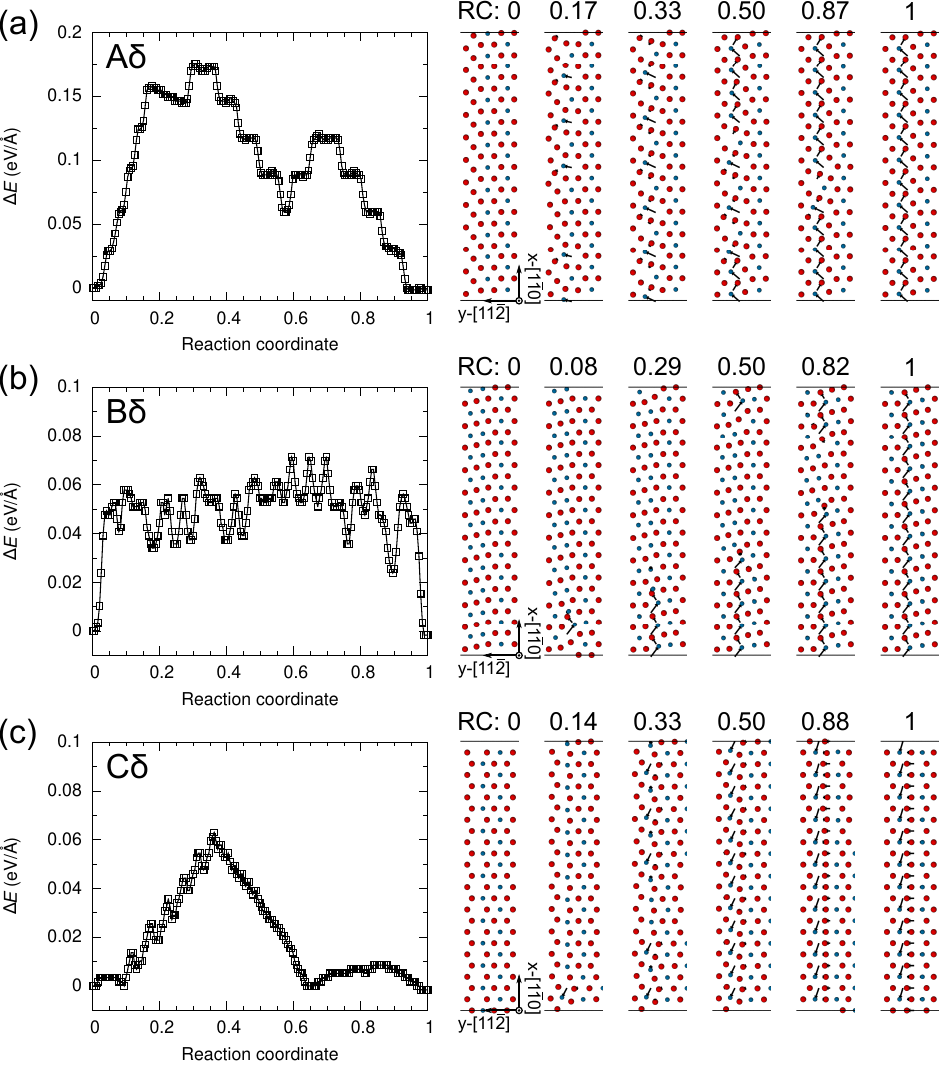}
    \caption{Transition mechanisms of the motion of synchro-Shockley partial dislocations in the \textit{C}15 NbCr$_{2}$ Laves phase with $l_x$ of 5.89\,nm: (a) 30\degree\ type II partial $A\delta$, (b) 30\degree\ type I partial $B \delta$, (c) 90\degree\ type I $C \delta$. Left: Excess energy as a function of reaction coordinate (RC), obtained from NEB calculations (note that the energy scale is larger in (a) than in (b) and (c)). Right: Mechanisms of synchro-Shockley partial dislocation motion associated with ISF extension, viewed along the slip-plane normal. Only atoms in the atomic layers where the dislocation glides are shown. Red (large) and blue (small) spheres represent Nb and Cr atoms, respectively. Black arrows indicate atomic displacement vectors.}
    \label{fig:partials}
\end{figure*}

\subsection{Motion of zonal dislocations in \textit{C}15 NbCr$_{2}$}\label{result:motion}

\subsubsection{Motion of synchro-Shockley partial dislocations} 

The mechanisms of motion of 30\degree\ type I and II  partials and 90\degree\ type I partial dislocations on the \hkl{111} plane in \textit{C}15 NbCr$_{2}$ and their associated energy profiles were analyzed using the NEB method. For the 30\degree\ type II partial $A\delta$, the transition events for motion were characterized as a non-sequential shuffling mechanism\cite{xie2023unveiling}. The MEP shows a continuously increasing energy profile associated with short-range displacements of Cr atoms and the accumulation of shear strain, followed by a subsequent energy drop corresponding to the rearrangement of Nb atoms as shown in Figure \ref{fig:partials}(a). The motion of 30\degree\ type I partial $B\delta$ proceeds via a kink-pair nucleation and propagation mechanism\cite{xie2023unveiling}, as illustrated in Figure \ref{fig:partials}(b). This process begins with a rate-limiting step involving the coupled activation of a kink-pair within the triple layers, followed by kink-pair propagation in opposite directions along the dislocation line via vacancy-hopping and interstitial-shuffling processes.  These localized mechanisms lower the overall energy barrier for dislocation glide by decomposing collective dislocation motion into localized atomic events, a behavior characteristic of crystals with high lattice friction \cite{caillard2003thermally}. The transition events and energy profiles of both 30\degree\ partials in \textit{C}15 NbCr$_{2}$ are consistent with previous atomistic simulations of \textit{C}14 CaMg$_{2}$ and \textit{C}15 CaAl$_{2}$ using modified EAM potentials \cite{xie2023unveiling,xie2023thermally}. For 90\degree\ type I partial $C\delta$, the transition events were also governed by a non-sequential shuffling mechanism similar to that of $A\delta^{30II}$, while the overall energy barrier is comparable to that of the $B\delta^{30I}$ partial. The MEP again exhibits a steadily increasing energy corresponding to Cr atom displacement and shear straining, followed by the rearrangement of Nb atoms, see Figure \ref{fig:partials}(c).

\subsubsection{Motion of coupled synchro-Shockley partial dislocations} 

\begin{figure*}[hbt!]
\centering
\includegraphics[width=\linewidth]{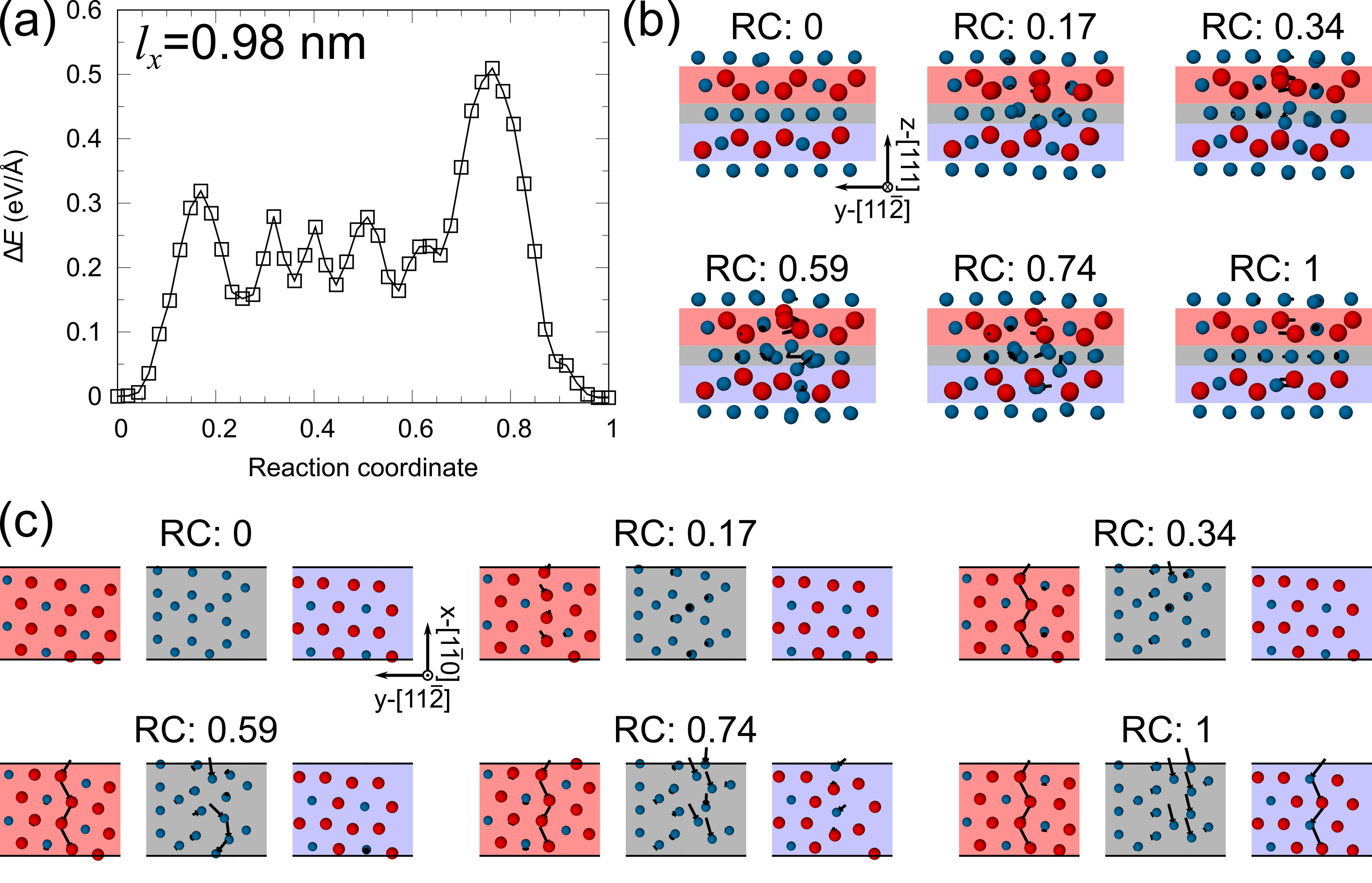}
\caption{Transition mechanism of ESF extension over one Peierls barrier via the motion of 90\degree\ type I and 30\degree\ type II synchro-Shockley partial dislocations in the \textit{C}15 NbCr$_{2}$ Laves phase with $l_x$ of 0.98\,nm. (a) Excess energy as a function of reaction coordinate (RC), obtained from NEB calculations. Mechanisms of zonal dislocation motion associated with ESF extension, viewed along (b) the \hkl[1-10] direction and (c) the slip-plane normal. Only atoms in the atomic layers where the dislocation glides are shown in (c). The triple layers where the 90\degree\ type I and 30\degree\ type II partials glide are highlighted with red and blue backgrounds, respectively, while the Kagomé layer between them is shaded in grey. Red (large) and blue (small) spheres represent Nb and Cr atoms, respectively. Black arrows indicate atomic displacement vectors.}
\label{fig:zonal_1peierls}
\end{figure*}

\begin{figure*}[hbt!]
\centering
\includegraphics[width=\linewidth]{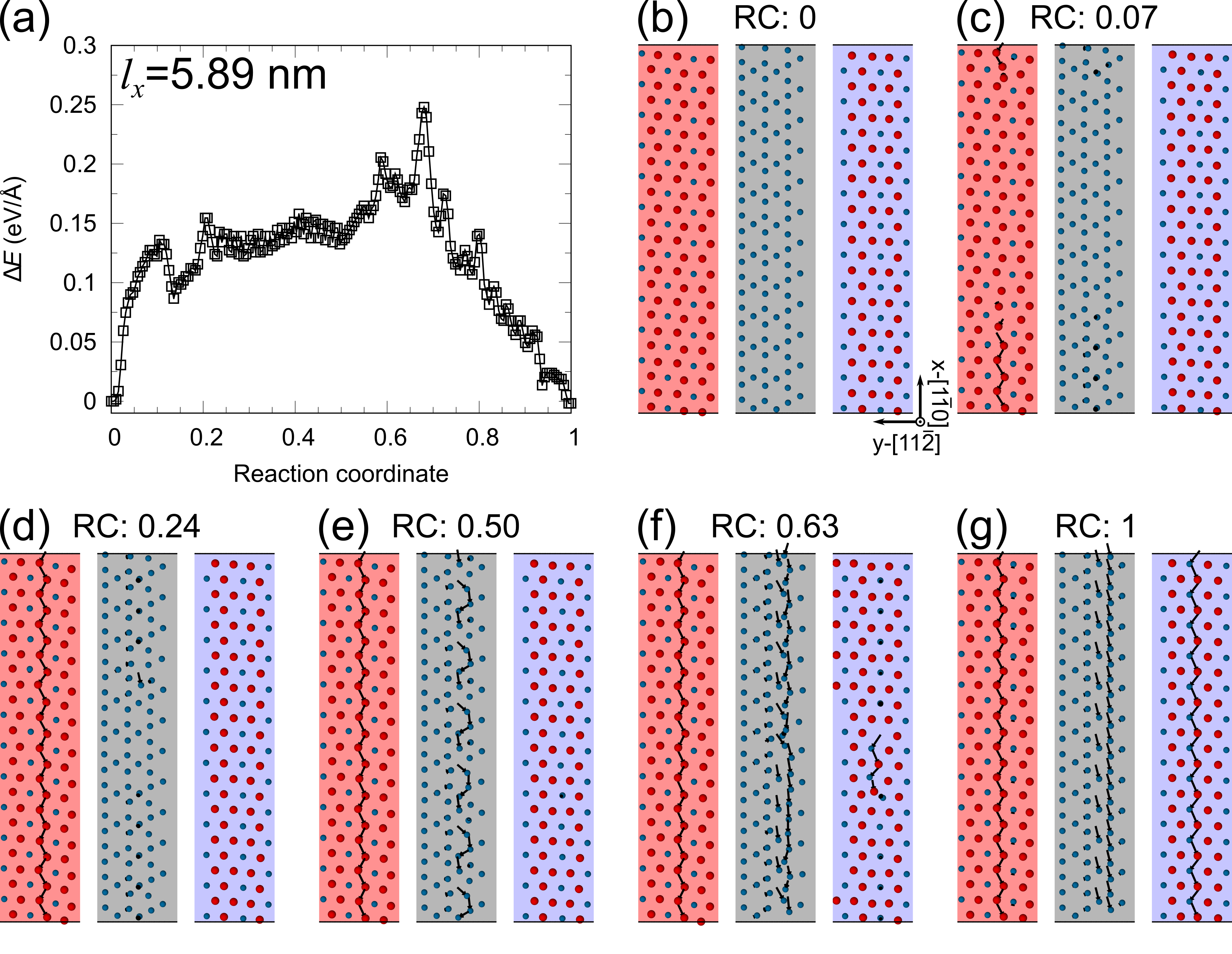}
\caption{Transition mechanism of ESF extension over one Peierls barrier via the motion of 90\degree\ type I and 30\degree\ type II synchro-Shockley partial dislocations in the \textit{C}15 NbCr$_{2}$ Laves phase with $l_x$ of 5.89\,nm. (a) Excess energy as a function of reaction coordinate (RC), obtained from NEB calculations. (b-g) Mechanisms of zonal dislocation motion associated with ESF extension, viewed along the slip-plane normal. Only atoms in the atomic layers where the dislocation glides are shown. The triple layers where the 90\degree\ and 30\degree\ partials glide are highlighted with red and blue backgrounds, respectively, while the Kagomé layer between them is shaded in grey. Red (large) and blue (small) spheres represent Nb and Cr atoms, respectively. Black arrows indicate atomic displacement vectors.}
\label{fig:zonal_1peierls_l59}
\end{figure*}

The ESF structure can be regarded as a combination of two ISFs, corresponding to the motion of two synchro-Shockley partials on adjacent \hkl{111} planes. Experimental observations in this study revealed the presence of two such partial dislocations (a 90\degree\ type I and a 30\degree\ type II synchro-Shockley partial), bounded by an ESF and stacked vertically with a merged core structure (Figure \ref{fig:exp_sim}). Additionally, these stacked partial dislocations were found to interact with synchro-Shockley partials on conjugate \hkl{111} planes, forming sessile locks at the intersections \cite{lockpaper}. The motion of the two synchro-Shockley partial dislocations gliding on adjacent \hkl{111} planes represents the fundamental mechanism governing ESF formation.

NEB calculations were performed to unveil the mechanism of motion of these two synchro-Shockley partial dislocations (a 90\degree\ type I and a 30\degree\ type II partials) or, equivalently, $\delta A$ \textit{or} $\delta B$ zonal dislocations bounded by an ESF in the \textit{C}15 NbCr$_{2}$ Laves phase. Dislocation lengths of $l_x$=0.98 and 5.89\,nm along the periodic direction, with approximately 95,000 and 570,000 atoms, respectively, were examined. For the $l_x$= 0.98\,nm setup, 48 images were used along the MEP, whereas the $l_x$= 5.89\,nm model, which is six times thicker, employed 192 images. For dislocation motion across a single Peierls barrier, results from both configurations consistently suggest that the two partials move sequentially rather than in a fully coupled manner. The MEP of ESF propagation shows multiple distinct transition events (see Figures~\ref{fig:zonal_1peierls} and \ref{fig:zonal_1peierls_l59}). As shown in the energy profile in Figure~\ref{fig:zonal_1peierls}(a), the first peak corresponds to the glide of the 90\degree\ type I synchro-Shockley partial on the upper triple layer. The subsequent peaks arise from atomic shuffling within the middle Kagomé layer between the two adjacent triple layers, while the final peak is associated with the glide of the 30\degree\ type II synchro-Shockley partial on the lower triple layer. 

The thicker dislocation setup ($l_x$=5.89\,nm) shows a similar energy profile and sequence of transition events, where the 30\degree\ type II partial on the lower triple layer glides following the motion of the 90\degree\ type I partial on the upper layer (Figure~\ref{fig:zonal_1peierls}). Specifically, the upper triple layer, where the 90\degree\ type I synchro-Shockley partial glides, extends the upper ISF via the kink-pair mechanism, which is absent during the motion of the isolated 90\degree\ type I partial (Figure~\ref{fig:zonal_1peierls_l59}(b-d)). Next, the middle Kagomé layer between the two adjacent triple layers undergoes partial deformation via atomic shuffling (Figure~\ref{fig:zonal_1peierls_l59}(d-f)). Finally, the lower triple layer, where the 30\degree\ type II partial glides, extends the lower ISF via another kink-pair process, accompanied by a cooperative rearrangement of the middle Kagomé layer (Figure~\ref{fig:zonal_1peierls_l59}(f-g)). These localized kink-pair processes are not fully captured in the thinner model, resulting in a higher overall energy barrier. Overall, these results demonstrate that the motion of the two synchro-Shockley partial dislocations is not strongly coupled when overcoming a single Peierls barrier.

\begin{figure*}[hbt!]
\centering
\includegraphics[width=\linewidth]{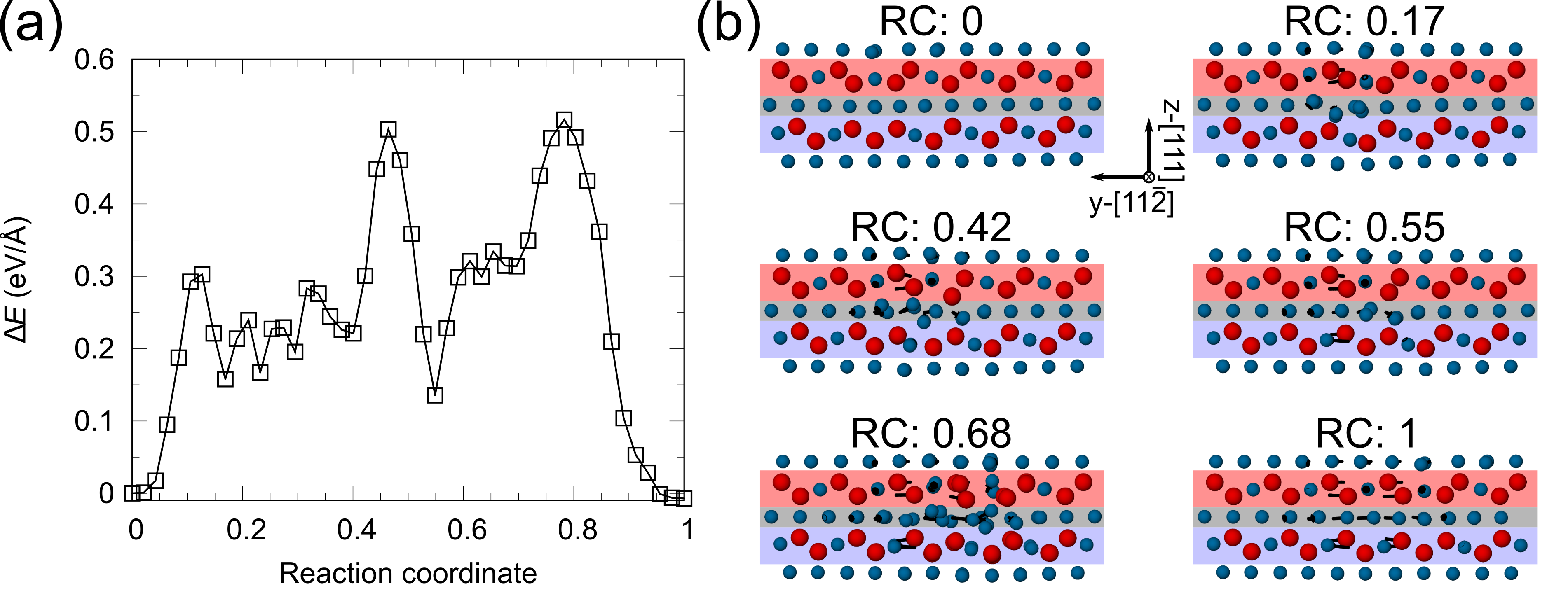}
\caption{Transition mechanism of ESF extension over two Peierls barriers via the motion of 90\degree\ type I and 30\degree\ type II synchro-Shockley partial dislocations in the \textit{C}15 NbCr$_{2}$ Laves phase with $l_x$ of 0.98\,nm. (a) Excess energy as a function of reaction coordinate (RC), obtained from NEB calculations. (b) Mechanisms of zonal dislocation motion associated with ESF extension, viewed along the \hkl[1-10] direction. The triple layers where the 90\degree\ type I and 30\degree\ type II partials glide are highlighted with red and blue backgrounds, respectively, while the Kagomé layer between them is shaded in gray. Red (large) and blue (small) spheres represent Nb and Cr atoms, respectively. Black arrows indicate atomic displacement vectors.}
\label{fig:zonal_2peierls}
\end{figure*}

Further NEB calculations of the motion of the coupled synchro-Shockley partial dislocations over a longer glide distance, spanning two Peierls barriers, were performed to elucidate the underlying mechanism. As demonstrated in the MEP analysis of ESF extension across a single Peierls barrier, the thinner dislocation setup ($l_x$= 0.98\,nm) is sufficient to capture the key transition mechanisms and sequence of events; therefore, this setup was adopted for investigating more complex reaction pathways. As shown in Figure~\ref{fig:zonal_2peierls}, the results reveal that the merged core structures are reproduced with a single Peierls valley periodicity, indicating that the upper 90\degree\ type I and the lower 30\degree\ type II partial tend to move in a coordinated manner that restores the stable merged core structure after traversing one Peierls barrier, following a similar mechanism as shown in Figure~\ref{fig:zonal_1peierls}. Although the two partials do not glide simultaneously into the next Peierls valley, the energetic preference for reestablishing the merged core structure suggests that they move alternately yet remain synchronized in their overall propagation along the slip plane. This alternating but coordinated motion defines the \textit{coupled synchro-shear slip mechanism}.

\begin{figure*}[hbt!]
\centering
\includegraphics[width=\linewidth]{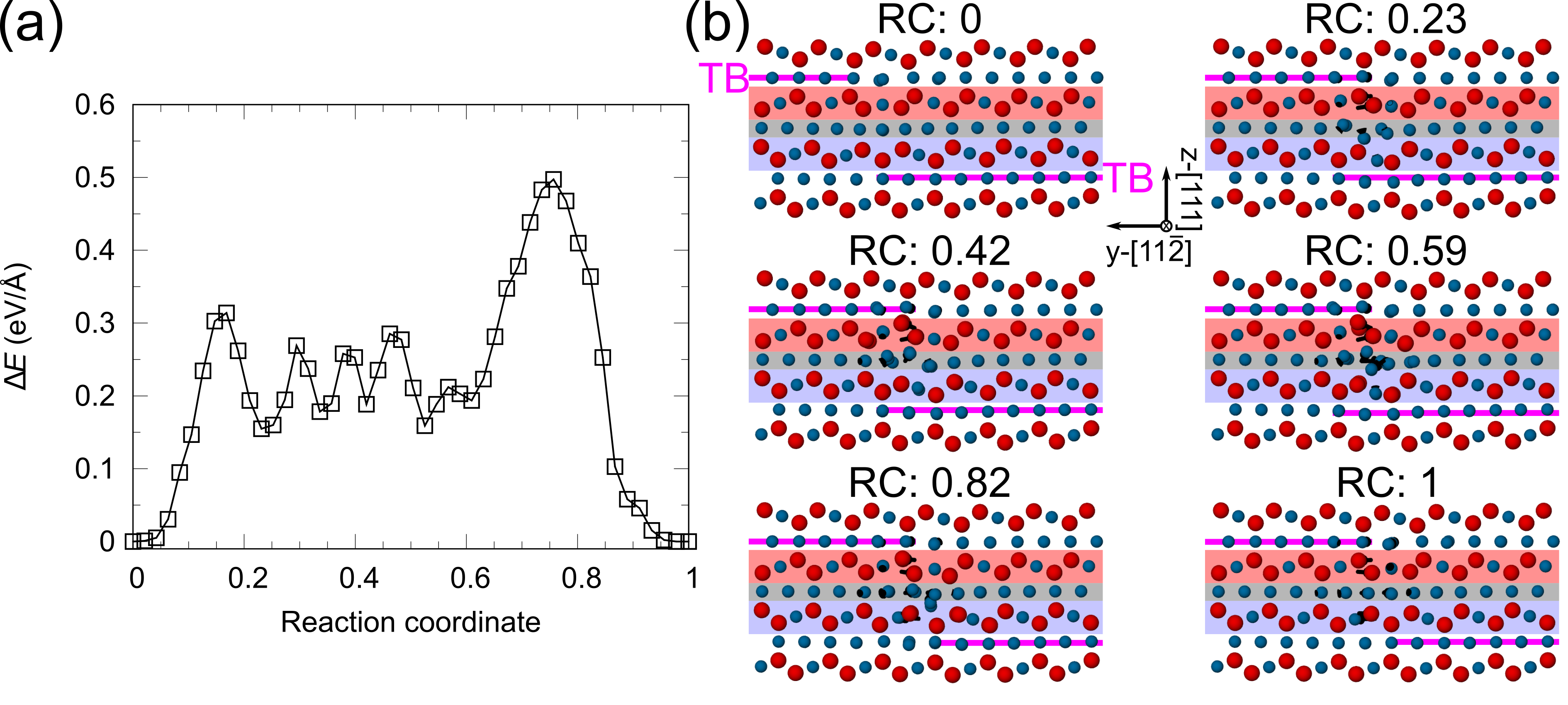}
\caption{Transition mechanism of twinning mediated via the motion of 90\degree\ type I and 30\degree\ type II synchro-Shockley partial dislocations in the \textit{C}15 NbCr$_{2}$ Laves phase with $l_x$ of 0.98\,nm. (a) Excess energy as a function of reaction coordinate (RC), obtained from NEB calculations. (b) Mechanisms of zonal dislocation motion adjacent to the twin boundary corresponding to the twinning process, viewed along the \hkl[1-10] direction. The triple layers where the 90\degree\ type I and 30\degree\ type II partials glide are highlighted with red and blue backgrounds, respectively, while the Kagomé layer between them is shaded in gray. Twin boundaries (TBs) are marked in magenta. Red (large) and blue (small) spheres represent Nb and Cr atoms, respectively. Black arrows indicate atomic displacement vectors.}
\label{fig:twin}
\end{figure*}

As evidenced by the direct observation of an ESF adjacent to a coherent twin boundary in the \textit{C}15 NbCr$_2$ Laves phase (see Figure~\ref{fig:exp_sim}(c)), twin growth can be interpreted as the consecutive propagation of ESFs. Understanding the coupled motion of two synchro-Shockley partial dislocations near a coherent twin boundary is essential for elucidating the mechanism of twin formation in Laves phases. Interestingly, the overall energy profiles and transition events for zonal dislocation motion in the \textit{C}15 matrix and near the twin boundary are nearly identical (Figure~\ref{fig:twin}), suggesting that the presence of the twin boundary does not significantly influence the energy barrier and mechanism for zonal dislocation motion. 

\subsubsection{Surface nucleation of coupled synchro-Shockley partial dislocations} 

\begin{figure*}[hbt!]
\centering
\includegraphics[width=\linewidth]{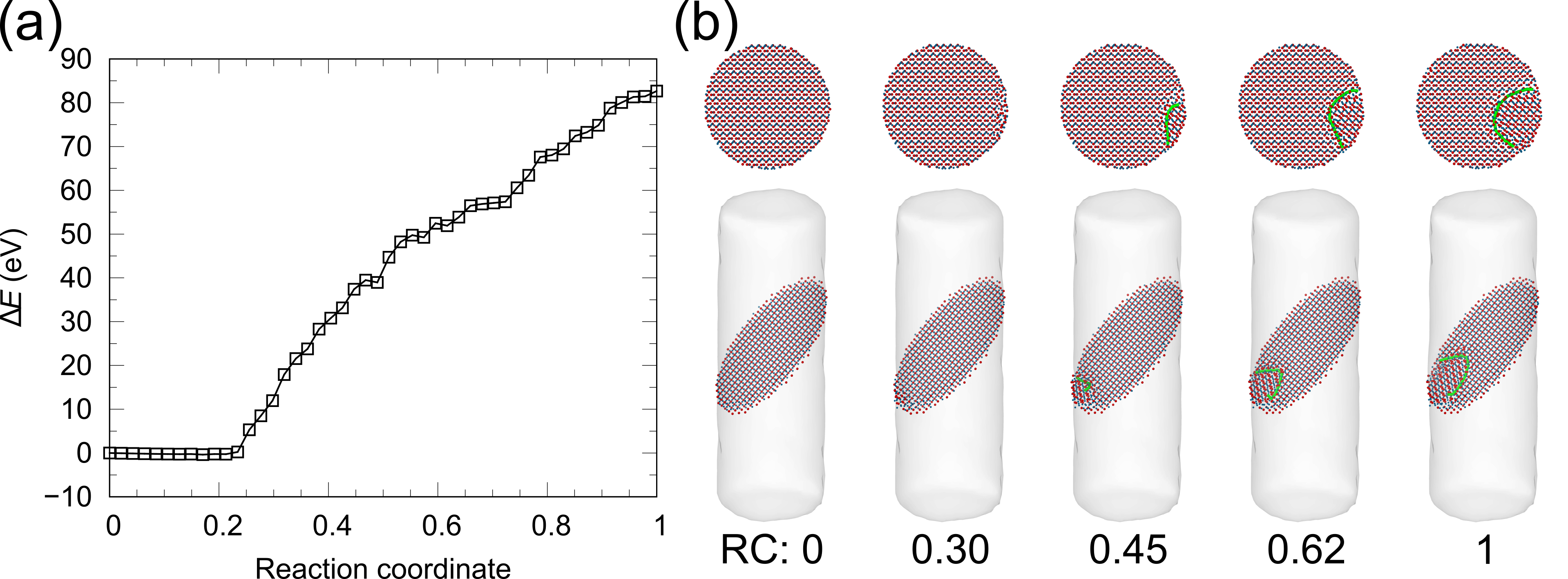}
\caption{Transition mechanisms of surface nucleation and propagation of coupled synchro-Shockley partial dislocations in a \textit{C}15 NbCr$_{2}$ nanopillar (diameter of 6.1\,nm and height of 18.3\,nm). (a) Excess energy as a function of reaction coordinate (RC), obtained from NEB calculations. (b) Mechanisms of zonal dislocation nucleation and propagation associated with ESF formation. Only atoms in the two adjacent triple layers where the dislocation glides are shown. Red (large) and blue (small) spheres represent Nb and Cr atoms, respectively. The zonal dislocation line is highlighted in green. Up: top view; down: perspective view with half-transparent surface mesh of the nanopillar.}
\label{fig:nucleation}
\end{figure*}

As demonstrated above, the coexistence and coupled motion of two synchro-Shockley partial dislocations are energetically favorable. To further investigate the heterogeneous nucleation of such a zonal dislocation, NEB calculations were performed. A nanopillar geometry with a circular cross-section (diameter of 6.1\,nm and height of 18.3\,nm) was employed to determine the MEP for dislocation nucleation and propagation, with both ends terminating at free surfaces. The initial configuration corresponds to a pristine nanopillar oriented with a \hkl(111) slip plane inclined at 45\degree\ to the pillar axis, while the final configuration contains an ESF bounded by a bowing-out zonal dislocation with a mixed character. The rigid-body motion of the nanopillar was suppressed during the calculations. The final state exhibits a higher energy than the pristine configuration due to the presence of the zonal dislocation bounded by the ESF, the formation of surface steps induced by dislocation glide, and the accumulation of strain energy (Figure~\ref{fig:nucleation}(a)). The intermediate states along the MEP reveal the coupled nucleation of two synchro-Shockley partial dislocations from the corner of the \hkl(111) slip plane, followed by their cooperative propagation along adjacent layers, see Figure~\ref{fig:nucleation}(b). Therefore, the heterogeneous nucleation of a zonal dislocation is energetically more favorable than the independent activation of isolated synchro-Shockley partials to form an ESF in the \textit{C}15 NbCr$_2$ Laves phase.

\section{Discussion} 

\subsection{Role of stacking fault energies in plasticity}\label{dis:SFE}

Stacking fault energies often play a central role in governing the characteristic deformation behavior and defect structures of crystalline materials. Both our atomistic simulations and previous DFT calculations \cite{hong2000first} consistently show that the energy of the ESF is lower than the energy of the ISF in the \textit{C}15 NbCr$_{2}$ Laves phase. This provides the thermodynamic driving force for the preferential formation of ESFs, explaining why ESFs are more frequently observed experimentally than isolated ISFs \cite{allen1972electron,chu1998phase}. The lower ESF energy in \textit{C}15 NbCr$_{2}$ also rationalizes the thermodynamic stability of the coupled synchro-Shockley partial dislocations bounded by an ESF observed in experiments, see Figure~\ref{fig:exp_sim}.

\begin{figure*}[ht!]
\centering
\includegraphics[width=\linewidth]{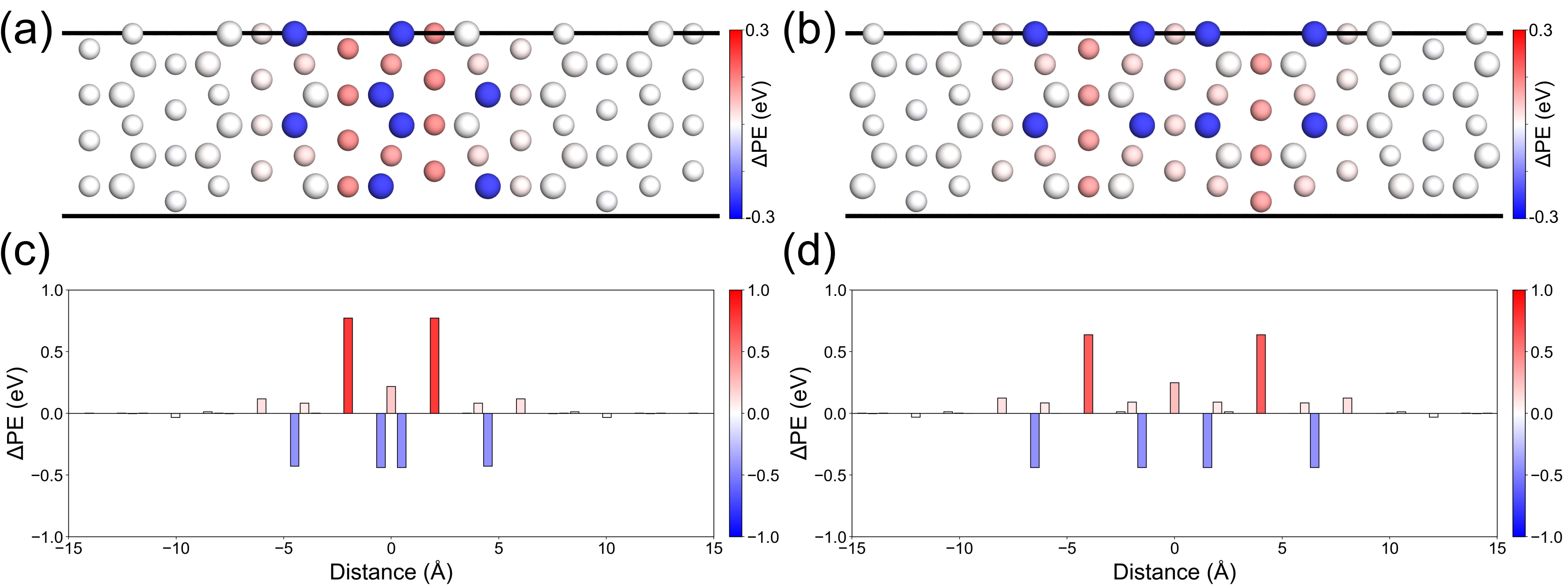}
\caption{Excess potential energies ($\Delta$PE) of (a) ISF and (b) ESF in the \textit{C}15 NbCr$_{2}$ Laves phase simulated using the ACE potential. Sum of excess potential energies of each \hkl(111) atomic layer in (c) ISF and (d) ESF.}
\label{fig:SFE}
\end{figure*}

Many ab-initio studies have employed an Ising-type model, based on the energy difference between various Laves phase polytypes ($\Delta \epsilon$), to estimate ISF and ESF energies in \textit{C}15 Laves phases \cite{chu1995stacking,sun2004ab,yao2007first}. For \textit{C}15 NbCr$_{2}$, this approach yields an ISF energy of approximately 90\,mJ/m$^{2}$ \cite{chu1995stacking}. However, this model considers only one and two stacking fault sequences for the ISF and ESF energy calculations, respectively, thereby accounting for only half of the stacking sequences involved in the actual faulted structures (see Figure~\ref{fig:SFE}). Correcting this error by doubling the fault energies predicted by the previous Ising model provides more realistic results:
\begin{equation}
    \gamma_\text{ISF} = \frac{48\Delta \epsilon_\text{\textit{C}14-\textit{C}15}}{\sqrt{3}a_0^2},
    \label{eq:ISF}
\end{equation}

\begin{equation}
    \gamma_\text{ESF} = \frac{96\Delta \epsilon_\text{\textit{C}36-\textit{C}15}}{\sqrt{3}a_0^2}.
    \label{eq:ESF}
\end{equation}

Within the present approximation, an ESF is treated as a thin twin embryo bounded by two equivalent TBs, with their mutual interaction neglected. The corresponding TB energy is:
\begin{equation}
    \gamma_\text{TB}=\frac{\gamma_\text{ESF}}{2}=\frac{48\Delta\epsilon_\text{\textit{C}36-\textit{C}15}}{\sqrt{3}a_0^2}.
    \label{eq:TB}
\end{equation}

For the ACE-simulated \textit{C}15 NbCr$_{2}$, the ISF energy predicted using the Ising-type model is 169.5\,mJ/m$^{2}$, which slightly overestimates the value from the supercell calculation (161.1\,mJ/m$^{2}$), while the predicted ESF energy is 147.2\,mJ/m$^{2}$, marginally lower than the reference value (148.8\,mJ/m$^{2}$). 
The predicted TB energy is 73.6\,mJ/m$^{2}$, compared with 74.4\,mJ/m$^{2}$ obtained directly from the supercell calculation. These discrepancies arise from this simplified Ising-type model, which assumes independent layer interactions and neglects the long-range elastic and electronic contributions inherent to the complex TCP lattice (see Figure~\ref{fig:SFE}).

Experimental estimation of SFE values has typically been attempted by measuring the radius of curvature of extended dislocation nodes or the splitting width of isolated dislocations \cite{dewit1959interaction,anderson2017theory}, e.g., by transmission electron microscopy. However, this line tension-based approach, commonly valid for FCC metals under near-equilibrium conditions, has produced inconsistent results for \textit{C}15 NbCr$_{2}$. Reported ISF energy values vary widely, from approximately 8\,mJ/m$^{2}$ to 60\,mJ/m$^{2}$ \cite{yoshida1994deformation,chu1995stacking,kazantzis1996stacking}, with the variation often attributed to temperature dependence \cite{kazantzis2007mechanical}. These experimental values are significantly lower than those predicted at 0 K by atomistic and ab-initio calculations, including 161.1\,mJ/m$^{2}$ from the ACE potential and 128-180\,mJ/m$^{2}$ from previous DFT studies \cite{chu1995stacking,vedmedenko2008first,liu2017first}. 

This large discrepancy reflects the kinetics-dominated nature of stacking fault states in Laves phases, rather than energetics, which arises from the high Peierls barrier and the thermally activated nature of dislocation motion \cite{xie2023thermally}. Synchro-Shockley partial dislocations are likely trapped within a Peierls potential valley rather than relaxing to their equilibrium positions. As a result, the experimentally measured splitting distances deviate substantially from their equilibrium values and cannot be directly correlated with the ISF energy. The frequent observation of coexisting \textit{C}14, \textit{C}15, and \textit{C}36 polytypes and their metastable intergrowths in \textit{C}15 NbCr$_2$ further supports that Laves phases can retain non-equilibrium polytypic layer structures following thermomechanical processing \cite{hajra2023high}, reflecting strong kinetic constraints that inhibit full relaxation toward thermodynamic equilibrium states. Overall, the influence of SFE on plasticity in Laves phases is significantly diminished by their high Peierls barrier and thermally activated nature, in sharp contrast to FCC metals, where the SFE directly governs dislocation and twinning behavior, as discussed later in Section~\ref{dis:twin}.

\subsection{Performance and validation of interatomic potentials}\label{dis:pot}

Accurate modeling of defect energetics and dynamics in Laves phases requires interatomic potentials that can capture the complex bonding characteristics in the TCP structures. In this study, the EAM\_Rösch potential was employed owing to its computational efficiency and demonstrated ability to reproduce key energetic and structural features of NbCr$_{2}$. Although our newly developed ACE potential offers higher fidelity in predicting energetics, their substantially greater computational cost limits their applicability for large-scale modeling, particularly in NEB calculations involving systems of several hundred thousand atoms and nearly two hundred intermediate images.

Comparative analyses of the MEPs for partial slip events obtained using the EAM\_Rösch and ACE potentials revealed excellent agreement in both activation energies and intermediate configurations along the slip trajectory (see Figure~\ref{fig:fault}). This consistency rationalizes the use of the EAM\_Rösch potential in this work, as it faithfully reproduces the energetics of slip processes while enabling efficient exploration of the large parameter space associated with dislocation reactions. Moreover, although the EAM\_Rösch potential predicts slightly negative SFE values, we believe this has an insignificant impact on the simulated dislocation behavior. As discussed above in Section~\ref{dis:SFE}, plasticity in Laves phases is controlled primarily by high Peierls barriers and thermally activated dislocation motion rather than by equilibrium stacking fault energetics. Consequently, the absolute magnitude of the SFE serves only as a secondary descriptor, and the close agreement of slip path energetics with the ACE potential results validates the reliability of the EAM\_Rösch potential for the present study.

In addition, the structures of dislocation cores and stacking faults obtained in simulations show excellent agreement with experimental observations, as shown in Figure~\ref{fig:exp_sim} and our previous study on dislocation locks involving zonal dislocation reactions \cite{lockpaper}. The samples prepared experimentally were heat-treated without applied deformation, indicating that the observed faults and dislocation structures are likely grown-in rather than deformation-induced defects. Nevertheless, these configurations provide valuable benchmarks for assessing atomistic models. The close match between experimentally observed and simulated core structures further confirms that the EAM\_Rösch and ACE potentials accurately capture the essential atomic-scale features governing defect stability in \textit{C}15 NbCr$_{2}$.

\subsection{Mechanisms of zonal dislocation motion}\label{dis:disloc}

\begin{figure*}[ht!]
\centering
\includegraphics[width=\linewidth]{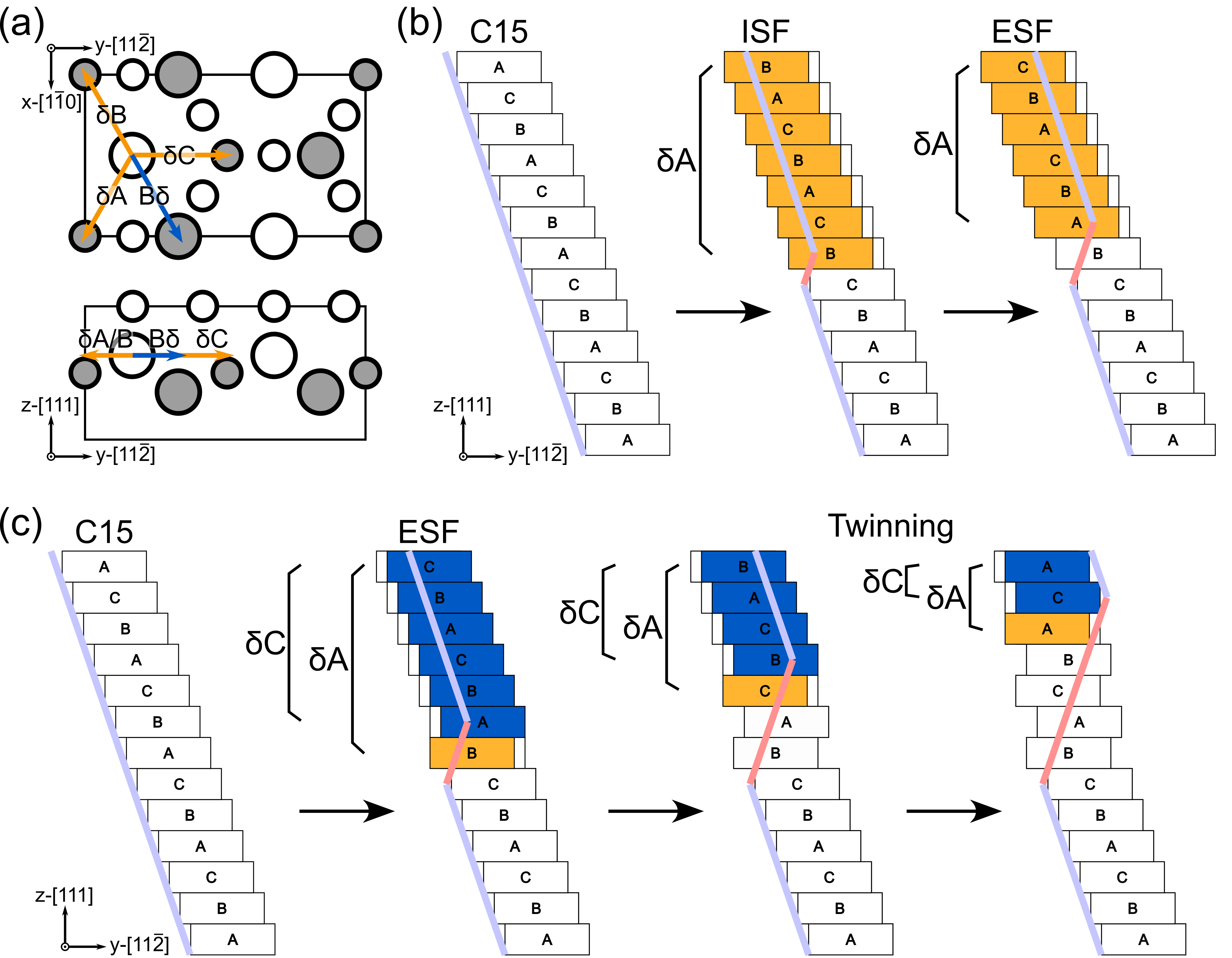}
\caption{Schematic illustration of (a) favorable slip systems, (b) conventional partial-dislocation-driven and (c) new zonal-dislocation-driven twinning routes in \textit{C}15 Laves phases. In (a), the favorable partial slip directions (Greek-Roman: $\delta A$, $\delta B$, and $\delta C$) on the $(111)$ plane in \textit{C}15 Laves phases are shown in blue, whereas the unfavorable Roman-Greek ones, like $B\delta$, are shown in red. Atoms in different layers along the $[111]$ direction are colored differently. Slip takes place in the upper atomic layers, colored in white.}
\label{fig:schematic}
\end{figure*}

From a crystallographic standpoint, the slip geometry of \textit{C}15 Laves phases differs from that of conventional FCC metals. In FCC metals, a \hkl{111} slip plane accommodates six equivalent partial slip directions, as the Greek–Roman and Roman–Greek partial slip paths are equivalent. In contrast, the \textit{C}15 Laves phase permits only three energetically favorable partial slip directions, corresponding exclusively to the Roman-Greek paths. This reduced symmetry arises from its interpenetrating sublattice structure, characterized by the alternating $\alpha$-c–$\beta$ atomic stacking within the triple layer where slip occurs (Figure~\ref{fig:schematic}(a)). For a dislocation line oriented along $\mathbf{x}$= \hkl[1-10], three distinct partial slip directions are geometrically and energetically favorable on the \hkl(111) slip plane, two 30\degree\ type I and II partials ($A\delta$ and $B\delta$) and one 90\degree\ type I partial ($C\delta$). In our previous work, the synchro-Shockley partial dislocations responsible for Lomer–Cottrell locks were identified as 30\degree\ partials \cite{lockpaper}. 

Earlier atomistic simulations of the \textit{C}14 CaMg$_{2}$ and \textit{C}15 CaAl$_{2}$ Laves phases using the modified EAM potentials revealed two distinct types of 30\degree\ synchro-Shockley partial dislocations with different core structures and mobilities \cite{xie2023unveiling,xie2023thermally}. These two partials can be regarded as leading and trailing components, bounding an ISF and representing the dissociation products of a perfect screw dislocation. In the present work, the same pair of 30\degree\ partials was also characterized in the simulated \textit{C}15 NbCr$_{2}$ Laves phase. Their core configurations and motion mechanisms closely resemble those reported in the \textit{C}14 structure, confirming the universality of these partials across different Laves polytypes. The 90\degree\ synchro-Shockley partial exhibits a significantly higher core energy compared with the 30\degree\ partials, explaining why the isolated 90\degree\ partial has not yet been observed experimentally in Laves phases.

At elevated temperatures, synchro-Shockley dislocations and associated ISF formation are widely recognized as the primary carriers of plastic deformation in Laves phases. As demonstrated in previous studies, thermal fluctuations are indispensable for their glide \cite{xie2023thermally}. In this work, NEB calculations further reveal that the propagation of ESFs proceeds through the cooperative motion of two synchro-Shockley dislocations, specifically a 90\degree\ type I and a 30\degree\ type II  partial ($C\delta$ + $A \delta$), via kink-pair processes. This coupled synchro-shear process is inherently thermally activated, with thermal fluctuations assisting the synchronized glide of both dislocations across adjacent atomic layers. The driving force for this coupled motion arises from the reduction of total core energy: the interaction between adjacent synchro-Shockley dislocations minimizes the local lattice distortion and stabilizes the faulted configuration relative to two isolated dislocations (see Figure~\ref{fig:paircore}).

Crystallographic orientation could play a critical role in determining whether deformation proceeds via coupled or single synchro-Shockley partials. The total Burgers vector of the coupled synchro-Shockley partials corresponds to $\delta B$. Orientation relationships that generate comparably high resolved shear stresses along both partial $A \delta$ and $C \delta$ directions promote the cooperative activation of coupled partials $\delta B$, enabling ESF propagation. In contrast, when the resolved shear stress is highest along a single partial slip direction, like $A \delta$ or $C \delta$, isolated synchro-Shockley partials are preferentially activated, resulting in ordinary slip and ISF formation. Overall, the propagation of ESFs via zonal dislocation motion represents an energetically favorable and kinetically accessible deformation mode, contributing to the plasticity of Laves phases by facilitating polytypic phase transformations and deformation twinning at elevated temperatures, as discussed later in Section~\ref{dis:twin}.

For the Nb-rich NbCr$_{2}$ Laves phase characterized experimentally in this work, Nb anti-sites and Cr vacancies were identified decorating dislocation cores and locks \cite{lockpaper}. The presence of point defect-decorated dislocation structures provides clear evidence of their diffusivity toward dislocations at elevated temperatures. The ability of point defects to migrate and accumulate at dislocation cores suggests that dislocation-point defect interactions may play a significant role in the deformation behavior of off-stoichiometric Laves phases, particularly under high-temperature conditions. Progressive softening behavior has been reported in Laves phases with deviations from stoichiometric compositions in experiments on \textit{C}15 NbCr$_{2}$ \cite{vignoul1993characterization,takasugi1996deformability,lu2009fracture} and other Laves phases \cite{voss2008composition,takata2016nanoindentation,luo2020composition,FREUND2024}, as well as computational studies \cite{xie2023unveiling,ladines2023off,FREUND2024}. This phenomenon has been attributed to the presence of structural point defects, such as antisites and vacancies, that lower the Peierls stress and thereby enhance dislocation mobility \cite{xie2023unveiling,FREUND2024}. Taken together, these observations suggest a possible contribution to dynamic strain aging, wherein dislocations interact with diffusing point defects, leading to serrated flow and stress anomalies, a phenomenon that warrants further detailed investigation.

\subsection{Zonal-dislocation-driven twinning mechanism}\label{dis:twin}

Laves phases exhibit twinning behavior that is fundamentally different from that of FCC metals. In FCC crystals, deformation twinning is carried out by the motion of leading Shockley partial dislocations (e.g., with a Burgers vector of $\delta A$ on the $(111)$ plane in Figure~\ref{fig:schematic}(b)) on successive \hkl{111} planes through the formation of ISFs. In contrast, twinning in cubic \textit{C}15 Laves phases is believed to occur through the cooperative motion of two coupled synchro-Shockley partials (e.g., $\delta C$ and $\delta A$ on the $(111)$ plane in Figure~\ref{fig:schematic}(c)) on adjacent planes, producing ESFs, as evidenced in our experiments (see Figure~\ref{fig:exp_sim}). This shear and coordinated atomic shuffling process, referred to as the zonal-dislocation-driven twinning mechanism, represents a more constrained multi-slip pathway than the conventional partial-dislocation-driven twinning widely observed in FCC metals. More specifically, the twinning process in Laves phases carried out by ESF formation requires the coupled activation of two co-planar partial slip systems ($\delta C$ and $\delta A$ on the $(111)$ plane), implying that twinning demands high and comparable resolved shear stresses along both slip directions, as also discussed above in Section~\ref{dis:disloc}. This unique behavior is consistent with experimental observations in high-temperature deformed \textit{C}15 NbCr$_{2}$, where grains oriented to facilitate simultaneous operation of two co-planar partial slip systems show a higher probability of twinning \cite{kazantzis2007mechanical}. These findings indicate that, rather than individual synchro-shear slip, the coupled synchro-Shockley partial motion, namely the zonal-dislocation-driven twinning mechanism, is a more plausible twinning route in cubic Laves phases.

In FCC metals, the competition between dislocation slip and twinning is primarily thermodynamically controlled by ISF energy \cite{gottstein2025materialwissenschaft,chen2003deformation}, owing to the low Peierls barrier for dislocation motion (on the order of a few to tens of MPa). A low ISF energy increases the separation between leading and trailing partials, promoting leading partial nucleation and thus favoring twinning. Conversely, a high SFE narrows the separation and promotes full dislocation activation. In contrast, deformation in Laves phases is kinetically controlled by the activation barrier associated with synchro-shear slip, which involves Peierls stresses orders of magnitude higher (a few GPa) than those in FCC metals. Therefore, the key energetic parameter governing plastic deformation in Laves phases is not the SFE itself, but the activation energy required for the motion of synchro-Shockley partials, as discussed above in Section~\ref{dis:SFE}. This barrier-controlled nature explains why \hkl{111} plasticity, including both dislocation slip and twinning, in Laves phases is promoted at elevated temperatures, where thermal activation facilitates the synchronized motion of partials.

The contrast in activation volumes further distinguishes the deformation behavior of FCC metals and Laves phase intermetallics. In FCC metals, full dislocation slip involves large activation volumes (100-1000\,b$^3$, as in Frank-Read source operation), while twinning, governed by the nucleation of leading partials, has much smaller activation volumes ($<$10\,b$^3$) and is therefore favored at low temperatures or high strain rates \cite{zhu2008temperature}. In contrast, in \textit{C}15 NbCr$_{2}$ Laves phase, high-temperature deformation experiments show the opposite trend: twinning becomes the primary deformation mode at high temperatures ($\geq$ 1350\,\degree C) or low strain rates ($\leq$ 1.75$\times$10$^{-4}$\,s$^{-1}$), while dislocation slip dominates at lower temperatures ($\leq$ 1300\,\degree C) or higher strain rates ($\geq$ 2.5$\times$10$^{-4}$\,s$^{-1}$) \cite{kazantzis2007mechanical}. This inversion arises because dislocation slip in Laves phases proceeds through local events such as kink-pair nucleation and propagation of synchro-Shockley partials, a process characterized by a small activation volume of the order of $\sim$10\,b$^3$ \cite{xie2023thermally}, consistent with experimental estimates \cite{kazantzis2007mechanical,kazantzis2008self}. Twinning occurs via the coupled activation of synchro-Shockley partials via two kink-pair processes and requires additional thermally activated atomic shuffle to coordinate the movement of adjacent planes, resulting in a somewhat larger activation volume. Experimentally, the activation volume for twinning in \textit{C}15 NbCr$_{2}$ has been estimated to be in the range of 15-80\,b$^3$ \cite{kazantzis2007mechanical,kazantzis2008self}. As a result, twinning in Laves phases becomes favorable only under conditions of high temperature or low strain rate, where sufficient thermal fluctuations are available to enable this cooperative motion. 
Consequently, the interplay between local crystallographic orientation and the relative activation barriers and volumes of single and coupled synchro-Shockley dislocations governs the transition between slip- and twin-dominated deformation modes, resulting in the reversed dependence of deformation mechanisms on temperature and strain rate in Laves phases compared with FCC metals. 

\section{Conclusions}
In this study, we elucidated a new slip mechanism, coupled synchro-shear slip, responsible for ESF propagation, thereby advancing the understanding of twinning and phase transformation mechanisms in Laves phases. This mechanism involves the alternating and synchronized glide of a 90\degree\ synchro-Shockley partial and a 30\degree\ synchro-Shockley partial via kink-pair nucleation and propagation, maintaining a stable merged core structure during dislocation motion. The finding of this coordinated slip process reveals that plastic deformation in Laves phases is inherently kinetic in nature, governed by thermally activated, barrier-controlled mechanisms rather than equilibrium fault energetics. This insight not only rationalizes the anomalous temperature- and strain-rate dependence of deformation mechanisms compared with FCC metals but also provides a framework for future investigations into phase transformations, alternative twinning pathways, and the dynamics of zonal and meta-dislocations in structurally complex intermetallic systems.

\section*{CRediT authorship contribution statement}
Sang-Hyeok Lee: Investigation, Methodology, Visualization, Formal analysis, Data curation, Writing – original draft, Writing – review \& editing.
Mariano Forti: Investigation, Methodology, Formal analysis, Data curation, Writing – review \& editing.
Thomas Hammerschmidt: Investigation, Methodology, Formal analysis, Data curation, Supervision, Funding acquisition, Writing – review \& editing.
Gang Liu: Investigation, Data curation, Writing – review \& editing.
Julien Guénolé: Methodology, Writing – review \& editing.
Siyuan Zhang: Investigation, Writing – review \& editing.
Gerhard Dehm: Funding acquisition, Writing – review \& editing, Resources, Supervision.
Sandra Korte-Kerzel: Funding acquisition, Project administration, Resources, Writing – review \& editing, Supervision.
Zhuocheng Xie: Investigation, Visualization, Funding acquisition, Writing – original draft, Writing – review \& editing, Supervision, Conceptualization.

\section*{Acknowledgments}
S.-H.L., S.Z., S.K.K., G.D., and Z.X. acknowledge financial support by the DFG through the SFB1394 Structural and Chemical Atomic Complexity – From Defect Phase Diagrams to Material Properties, project ID 409476157. S.K.K. and Z.X. are grateful for funding from the European Research Council (ERC) under the European Union’s Horizon 2020 research and innovation programme (grant agreement No. 101168203 TailorPlast). Z.X. acknowledges financial support by the DFG – project ID 562592407.
The authors gratefully acknowledge the computing time provided to them at the NHR Center NHR4CES at RWTH Aachen University (project number p0021574). This is funded by the Federal Ministry of Education and Research, and the state governments participating on the basis of the resolutions of the GWK for national high performance computing at universities (www.nhr-verein.de/unsere-partner). The data used in this publication were managed using the research data management platform Coscine with storage space granted by the Research Data Storage (RDS) of the DFG and Ministry of Culture and Science of the State of North Rhine-Westphalia (DFG: INST222/1261-1 and MKW: 214-4.06.05.08 - 139057). The authors would like to thank Johannes Roth (Universität Stuttgart) for providing the Nb-Cr EAM interatomic potential.

\bibliographystyle{elsarticle-num}
\bibliography{main}

\end{document}